\documentclass[useAMS,usenatbib]{mn2e}
\usepackage[dvips]{graphicx}
\usepackage{amssymb}

\newcommand{\kms}{\rm km\,s$^{-1}$}

\begin{document}

\title[Mid-IR Spectral Properties of IR~QSOs]{Mid-Infrared Spectroscopic Properties of Ultra-Luminous Infrared Quasars}
\author[C.Cao et al.]{
Chen Cao,$^{1,2,3,4}$\thanks{E-mail:~{\tt ccao00@gmail.com}}
X. Y. Xia,$^{5}$\thanks{E-mail:~{\tt xyxia@bao.ac.cn}}
Hong Wu,$^{3}$
S. Mao,$^{6,5}$
C. N. Hao,$^{7,5}$
Z. G. Deng$^{4,5,3}$\\
$^{1}$School of Space Science and Physics, Shandong University at Weihai, Weihai, Shandong 264209, China\\
$^{2}$Visiting Scholar, Harvard-Smithsonian Center for Astrophysics, Cambridge, MA 02138\\
$^{3}$National Astronomical Observatories, Chinese Academy of Sciences, Beijing 100012, China\\
$^{4}$Graduate University, Chinese Academy of Sciences, Beijing 100039, China\\
$^{5}$Tianjin Astrophysics Center, Tianjin Normal University, Tianjin 300384, China\\
$^{6}$Jodrell Bank Centre for Astrophysics, Alan Turing Building, University of Manchester, Manchester M13 9PL, UK\\
$^{7}$Institute of Astronomy, University of Cambridge, Madingley Road, Cambridge CB3 0HA, UK\\
}
\maketitle

\label{firstpage}

\begin{abstract}
We analyse mid-infrared (MIR) spectroscopic properties for 19 ultra-luminous infrared quasars (IR~QSOs) 
in the local universe based on the spectra from the Infrared Spectrograph on board the {\it Spitzer Space 
Telescope}. The MIR properties of IR~QSOs are compared with those of optically-selected Palomar-Green 
QSOs (PG~QSOs) and ultra-luminous infrared galaxies (ULIRGs). The average MIR spectral features from 
$\sim$5 to 30$\mu$m, including the spectral slopes, 6.2$\mu$m PAH emission strengths and [NeII] 12.81$\mu$m 
luminosities of IR~QSOs, differ from those of PG~QSOs. In contrast, IR~QSOs and ULIRGs have comparable 
PAH and [NeII] luminosities.  These results are consistent with IR~QSOs being at a transitional stage 
from ULIRGs to classical QSOs. We also find that the colour index $\alpha$(30, 15) is a good indicator of the 
relative contribution of starbursts to AGNs for all QSOs. Correlations between the [NeII] 12.81$\mu$m and 
PAH 6.2$\mu$m luminosities and those between the [NeII], PAH with 60$\mu$m luminosities for ULIRGs and IR~QSOs 
indicate that both [NeII] and PAH luminosities are approximate star formation rate indicators for IR~QSOs 
and starburst-dominated galaxies; the scatters are, however, quite large ($\sim$ 0.7 to 0.8~dex). Finally 
the correlation between the EW (PAH 6.2$\mu$m) and outflow velocities suggests that star formation activities 
are suppressed by feedback from AGNs and/or supernovae.
\end{abstract}

\section{Introduction}
Since the discovery of ultra-luminous infrared galaxies (ULIRGs, L$_{\rm IR}$$>$10$^{12}L_{\odot}$) by the 
{\it Infrared Astronomical Satellite} ({\it IRAS}) in the 1980's \citep[e.g.,][]{Houck85}, it is widely 
accepted that ULIRGs result from strong interactions/mergers between gas-rich disk galaxies. These 
mergers form elliptical galaxies and ULIRGs are an important intermediate stage in the process during which 
at least a fraction of ULIRGs manifest as dust-enshrouded QSOs \citep[see, e.g.,][]{Sanders96, Lonsdale06}. 
In addition, active galactic nuclei (AGNs) triggered by mergers tend to appear at the final merging stages 
(e.g., \citealt{Sanders88}, \citealt{Zheng99}, \citealt{Cui01}, \citealt{Veilleux02} and reference therein).

There is mounting evidence that QSOs with far-infrared (FIR) excess have massive starbursts in their 
host galaxies. For example, \citet{CS01} investigated 9 QSOs with FIR excess and found that their host 
galaxies are tidally interacting or major merger systems with obvious recent star-forming activities. 
From the detections of mid-infrared (MIR)/FIR\footnote{In this paper, MIR refers to 5-35 $\mu$m, FIR 
35-350 $\mu$m, and IR 8-1000$\mu$m.} emissions for FeLoBALs (Broad Absorption Line QSOs with 
low-ionisation and iron absorption lines) by Multiband Imaging Photometer on {\it Spitzer} 
\citep[MIPS;][]{Rieke04} on board the {\it Spitzer Space Telescope} \citep{Werner04}, \citet{Farrah07a} 
find that all their 9 FeLoBALs are extremely infrared (IR) bright, and concluded that these QSOs 
are in transition from ULIRGs to classical QSOs with ongoing or recent starbursts, because the iron 
absorption may be from iron ejected during starbursts. \citet{Hao05} studied 31 QSOs/Seyfert'1s selected 
from the local ULIRG samples (termed as IR~QSOs for simplicity). By comparing the FIR spectral index of 
IR~QSOs with those of optically selected Palomar-Green QSOs (PG~QSOs; \citealt{Schmidt83}), they argued 
that the FIR excess of IR~QSOs relative to PG~QSOs is from massive starbursts and inferred star formation 
rate (SFR) in the host galaxies of IR~QSOs. Recently, from studies of z$\sim$6 QSOs with strong sub-mm 
emissions, \citet{Carilli07}, \citet{WangR07} and \citet{WangR08} concluded that massive starbursts also 
exist in their host galaxies. The conclusion is consistent with the results of \citet{Hao08} that high redshift 
(sub)mm-loud QSOs follow the same trend for FIR to bolometric luminosities established by low redshift IR~QSOs 
\citep{Hao05}. All these studies suggest that there exists a transitional stage, during which both the central 
black hole and the spheroidal component of QSO hosts grow rapidly in a coeval fashion.

However, there is still a debate about the origin of FIR emission from QSOs, because one cannot firmly exclude 
the possibility that FIR emissions are from dust tori heated by central AGNs \citep[for detailed discussions 
see][]{Haas03}. Moreover, from the molecular gas properties of PG~QSOs with IR excess and comparisons with ULIRGs, 
\citet{Evans06} find that the $L_{\rm IR} / L'_{\rm CO}$ and $L_{\rm IR} / L'_{\rm HCN}$ ratios for PG~QSOs are higher 
than those of ULIRGs, implying that AGNs contribute significantly to the dust heating and hence to the FIR emission. 
Therefore, other SFR indicators besides the FIR emission for QSOs are important for further understanding the 
coeval growth of supermassive black holes and their host galaxies.

Recently, the QUEST (Quasar and ULIRG Evolution Study) group \citep[see][]{Schweitzer06, Netzer07} reported the 
detection of polycyclic aromatic hydrocarbon (PAH) emission features in PG~QSOs using the Infrared Spectrograph 
(IRS) on {\it Spitzer} \citep{Houck04}. For 11 out of 26 PG~QSOs PAHs have been clearly detected. Furthermore, 
the average spectrum of the undetected 15 PG~QSOs also shows PAH features. Since the PAH emissions are closely 
related to star formation, not to AGNs \citep[see, e.g.,][]{Shi07}, such detections strongly suggest that star 
formation occurs widely in QSOs. Their analysis shows that 30\% or more FIR emission in these PG~QSOs is from 
starbursts. Furthermore, given that the low-excitation fine-structure emission line [NeII] 12.81$\mu$m is one 
of the dominant emission lines of HII regions and that the PAH molecules are easily destroyed by high energy 
photons from AGNs \citep[e.g.,][]{Wu07}, [NeII] emission may be an alternative, perhaps even better, tracer 
of star formation for QSOs (see \S4).

In this paper we study the MIR spectroscopic properties of IR~QSOs based on {\it Spitzer} IRS observations, and 
examine their connections and evolutionary relations to ULIRGs and PG~QSOs. The sample selection, data acquisition 
and reduction are described in Sect. 2 and 3. The major results and discussions are given in Sect. 4 and 5. Finally 
we summarise our results in Sect. 6. We adopt cosmological parameters {\it H}$_{\rm 0}$=70 km s$^{-1}$ Mpc$^{-1}$, 
$\Omega$$_{\rm m}$=0.3, and $\Omega$$_{\rm \Lambda}$=0.7 throughout this paper.

\section{Sample Selection} \label{sec:sample}
IR~QSOs are defined as type 1 AGN with L$_{\rm IR}$($8-1000 \mu$m)\,$>$\,10$^{12}$\,L$_{\odot}$ \citep{Zheng02}. 
Our basic IR~QSO samples are compiled from ULIRG samples with spectroscopic information, plus the IR~QSOs 
obtained directly from the cross-correlation of the {\it IRAS} Point-Source catalog with the {\it ROSAT} 
All-Sky Survey Catalog. The ULIRG samples consist of 118 ULIRGs from 1$\,$Jy ULIRGs survey \citep{Kim98} 
and 97 ULIRGs from the QDOT redshift survey \citep{Lawrence99}. The total number of IR~QSOs is 31, about 
one third of all the IR~QSOs found in a complete redshift survey with 15,411 {\it IRAS} galaxies and about 
900 ULIRGs (PSCz; \citealt{Saunders00}). Thus it should be a representative sample of IR~QSOs (see 
\citealt{Zheng02} and \citealt{Hao05} for more detailed descriptions).

We searched the {\it Spitzer} archival data and found that 18 out of 31 IR~QSOs have been observed by IRS and the 
data are available (see Table~1). Notice that 9 of the 10 IR~QSOs (out of a total of 118 ULIRGs) from the 1$\,$Jy 
ULIRG sample are included in our sample. The other 9 IR~QSOs are from QDOT (4) and other QSO samples. In addition 
we include the object IRAS~F14026+4341, which is classified as a hyper-luminous infrared galaxy (with 
L$_{\rm FIR}$ $>$ 10$^{13}$L$_{\odot}$, \citealt{Rowan-Robinson00}) and a broad absorption line quasar \citep{Low89}. 
Our sample includes 90\% (50\%) IR QSOs out of 1$\,$Jy (QDOT) ULIRGs, and thus should be an overall representative 
sample. We checked that our results are essentially unchanged if we focus only on the 9 IR QSOs from the 1$\,$Jy 
ULIRG samples, and thus our compiled sample has no significant biases.

14 of the 19 objects have both low- and high-resolution IRS observations, while four (3C~48, IRAS~F02054+0835, 
PG~1543+489, and IRAS~F20036$-$1547) have only low-resolution observations, and one (IRAS~F21219$-$1757) has no 
Long-Low (LL) mode (see $\S$3.1) observation (see Table~3).

The IRS low-resolution spectra of a sample of Palomar-Green QSOs (PG~QSOs) are retrieved from {\it Spitzer} GTO 
and GO archival data (programs 14, 3187, 3421, and 20142; see Table~4). We remove objects whose MIR spectra have 
a low S/N ratio or redshift larger than 0.27 to guarantee reliable measurements of rest-frame 30$\mu$m fluxes. 
The number of PG~QSOs is 19 (see Table~1), the same as the number of IR~QSOs. This PG~QSO sample is mainly used 
for studying the MIR spectral slopes benefiting from the full wavelength coverage from $\sim$5-30$\mu$m of their 
low-resolution spectra. We also collected another PG~QSO sample with 22 objects studied 
by \citet{Schweitzer06} (after excluding four objects that have been classified as IR~QSOs and grouped
into IR~QSO sample). This second PG~QSO sample has deep SL mode observations (5.2-14.5$\mu$m)
and is thus suitable for studying weak PAH emission features in continuum-dominated 
QSOs, and for analysing properties of the MIR fine-structure lines (especially [NeII]12.81$\mu$m line in this work) 
from high-resolution observations. 

The sample of Ultra-Luminous Infrared Galaxies (ULIRGs) is selected based on the {\it IRAS} 1~Jy sample of ULIRGs 
\citep{Kim98}, which have optical spectroscopic observations by \citet{Veilleux99} and \citet{Wu98}. Their IRS 
low-resolution spectra are retrieved from {\it Spitzer} GTO archival data (program ID 105; see Table~4). The number 
in our ULIRG sample is 35 (see Table~2), including all spectral types except type 1 AGNs, namely Seyfert~2's, LINERs, 
and HIIs as classified by \citet{Veilleux99} and \citet{Wu98} from diagnostic diagrams of optical lines.

\section{Data Acquisition and Reduction} \label{sec:data}

\setcounter{figure}{0}
\begin{figure*}
\includegraphics[angle=90,scale=.65]{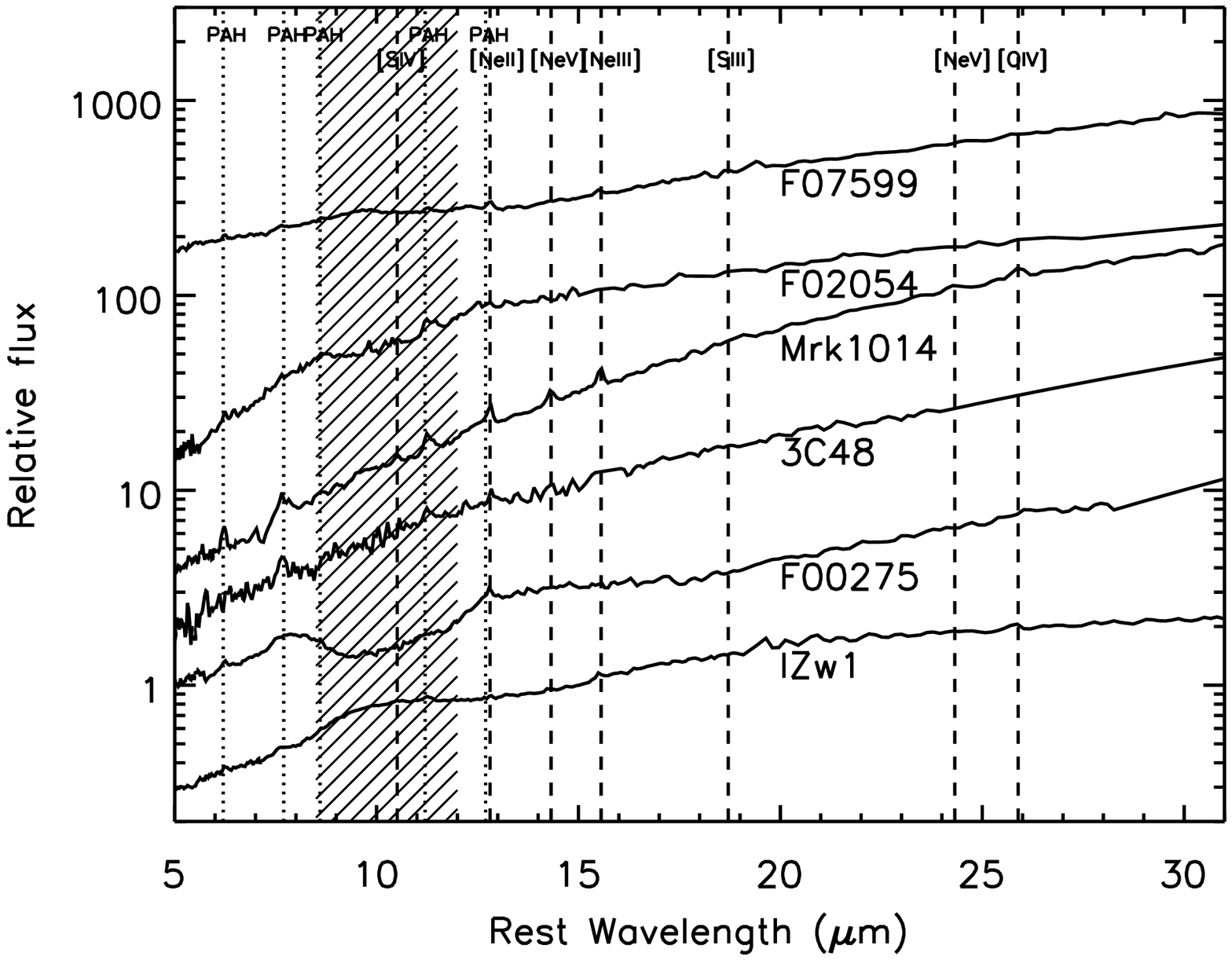}
\hfill
\includegraphics[angle=90,scale=.65]{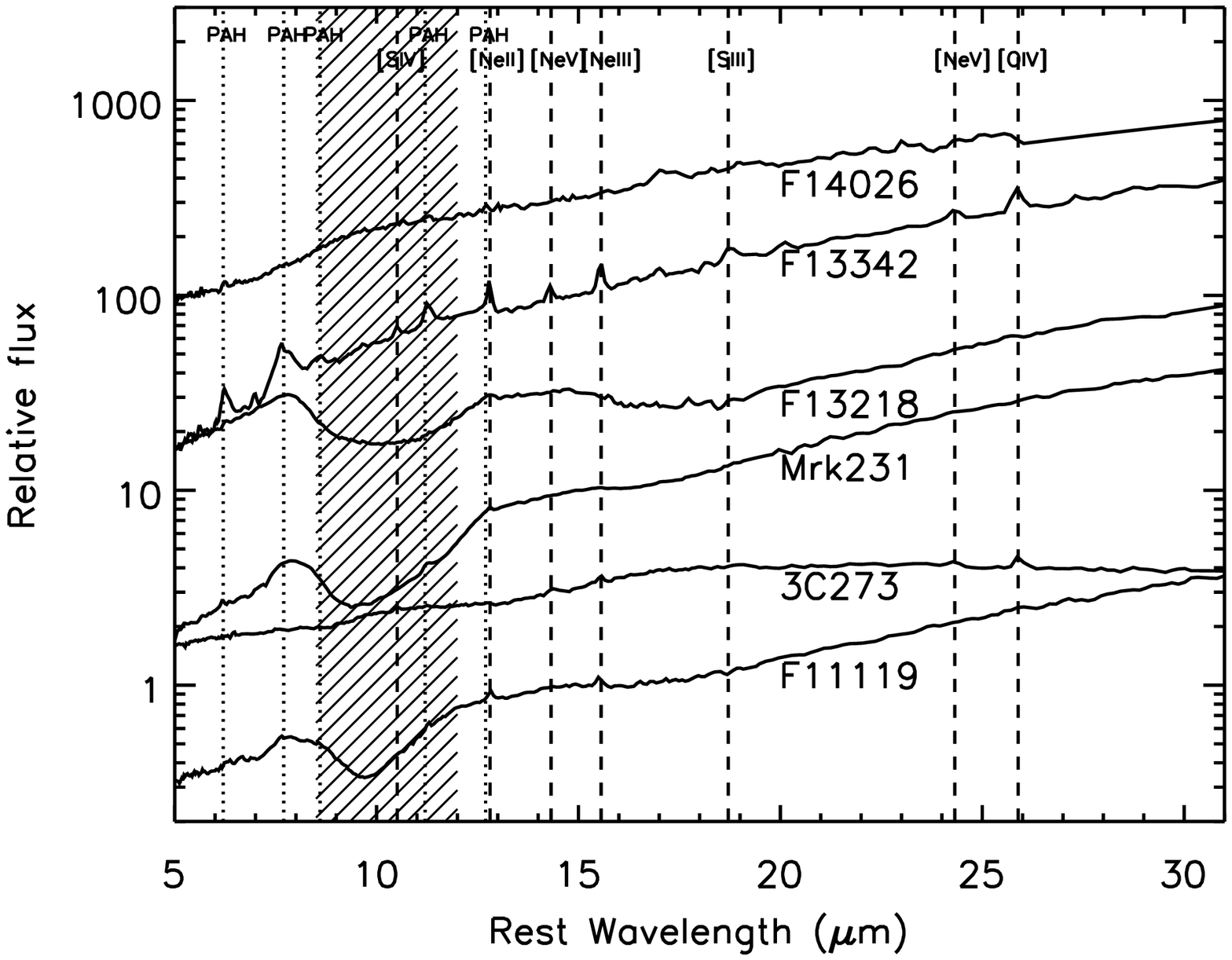}
\caption{Low-resolution {\it Spitzer}/IRS mid-infrared spectra of 18 IR~QSOs in our sample (except for IRAS~F21219$-$1757 
which has no LL mode observation), all have been de-redshifted and shifted upward for clarity. The dotted lines show 
the PAH features at 6.2, 7.7, 8.6, 11.2 \& 12.7 $\mu$m, and the dashed lines show the mid-infrared fine-structure 
lines of [SIV]10.51$\mu$m, [NeII]12.81$\mu$m, [NeV]14.32$\mu$m, [NeIII]15.56$\mu$m, [SIII]18.71$\mu$m, [NeV]24.32$\mu$m 
\& [OIV]25.89$\mu$m. The shaded bar denotes the silicate emission/absorption feature centred at 9.7$\mu$m.} 
\label{fig_spec}
\end{figure*}
\setcounter{figure}{0}
\begin{figure*}
\includegraphics[angle=90,scale=.65]{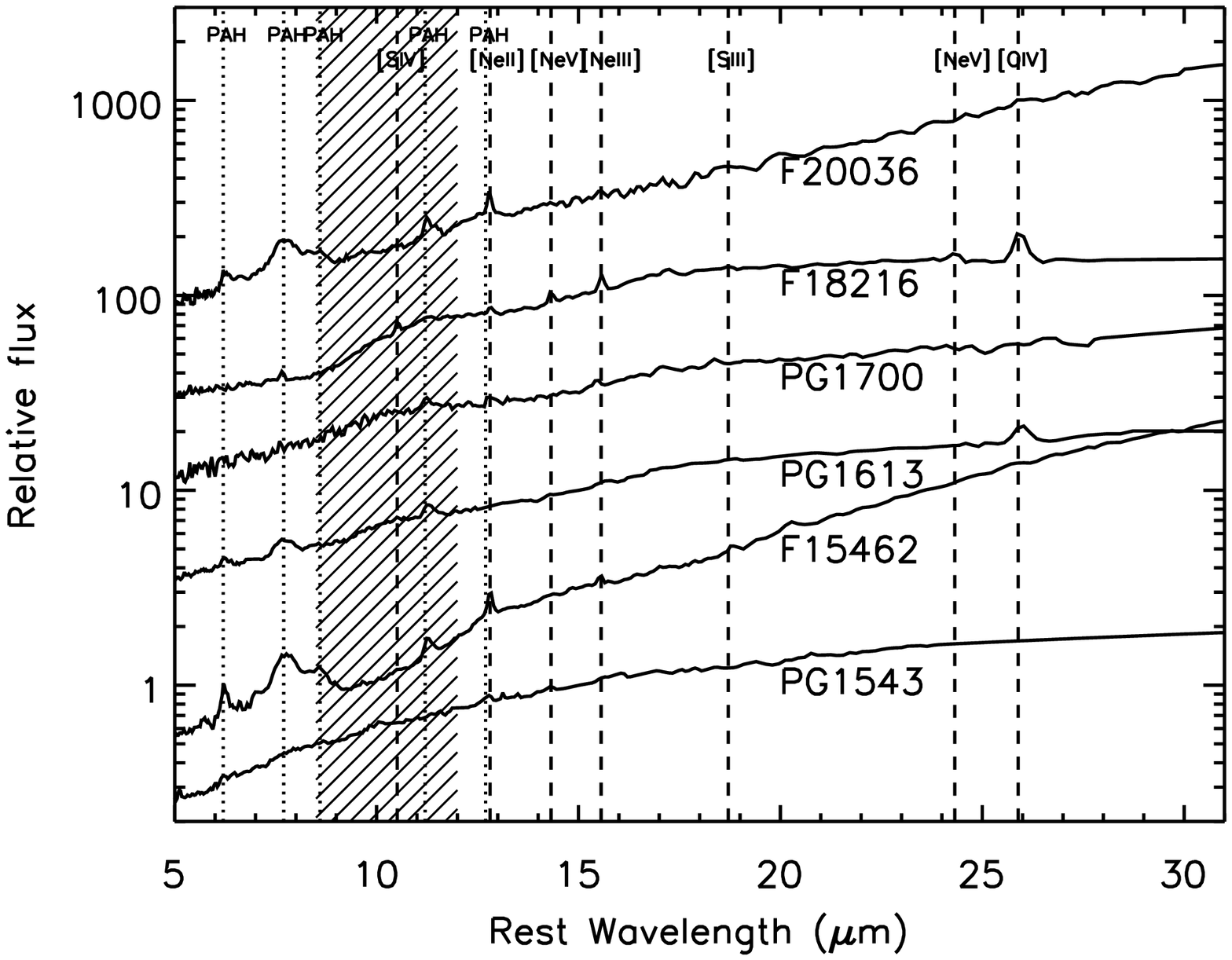}
\caption{Continued.}
\end{figure*}

\subsection{Mid-infrared spectra from {\it Spitzer} IRS}

The MIR spectra are acquired from the {\it Spitzer} archival data using the Leopard software (see Table~3 for the 
integrated exposure times and program IDs for IR~QSOs). The data (versions 13.2 to 15.3 of {\it Spitzer} pipeline 
reduction) include low-resolution (Short-Low [SL] \& Long-Low [LL] modes, R$\sim$60-120 and wavelength range: 
5.2-38.0$\mu$m) IRS spectra for IR~QSOs, PG~QSOs, and ULIRGs, and high-resolution (Short-High [SH] \& Long-High [LH] 
modes, R$\sim$600 and wavelength range: 9.9-37.2$\mu$m) IRS spectra for most of the IR~QSOs. We use the SMART software 
\citep{Higdon04} for data reduction, including the removal of rogue pixels, sky subtraction, and spectral extraction and 
analysis. The sky backgrounds for low-resolution (SL \& LL modes) spectra are subtracted by differencing the adjacent 
sub-slit positions (1st \& 2nd nods). No background subtraction is performed for the high-resolution (SH \& LH modes) 
spectra, but this does not affect MIR fine-structure line measurements \citep[see, e.g.,][]{Farrah07b}. The slit widths 
of 3$''$.6 to 11$''$.1 include most of the emission from the QSO and its host galaxy, so no aperture corrections 
are performed. For the low-resolution spectra we use the 12$\mu$m and 25$\mu$m flux densities from {\it IRAS} (or 
{\it ISO} if the {\it IRAS} fluxes are upper limits or not available) to scale the spectra by multiplying by a small 
factor, which is more suitable for the comparison between MIR and FIR properties in our statistics. The scaling 
factors for IR~QSOs and ULIRGs are often close to unity, typically less than 1.1\footnote{One exception is 3C~273 
for which no scaling was performed since it exhibits large variabilities in the MIR \citep{Neugebauer99, HL05}.}. 
However, for PG~QSOs the scaling factors are larger, typically $\sim$1.5. This may be caused by variabilities 
of quasars in MIR \citep[e.g.,][]{Neugebauer99} and/or contamination of companions (environments) for {\it IRAS} 
(or {\it ISO}) measurements. 

\subsection{Measurements of PAH and mid-infrared fine-structure lines}

The fluxes of PAH emission at 6.2$\mu$m are measured by integrating the flux above a local continuum from 6.0-6.5$\mu$m 
approximated by a second-order polynomial \citep[e.g.,][]{Spoon07, Desai07}. The uncertainties (1$\sigma$) in the 
measurements are 20$\%$ on average (varying from $\sim$5$\%$ for PAH strong objects to about 50$\%$ for those with 
only marginally detectable PAH features). The equivalent widths (EW) of 6.2$\mu$m PAH feature are obtained from dividing 
the integrated PAH flux by the continuum flux density below the peak of the feature. Upper limits (3$\sigma$) are 
given by adopting typical widths of $\sim$0.2$\mu$m for the 6.2$\mu$m PAH feature \citep{Smith07}, which is similar 
to the value, $\sim$0.6$\mu$m, used for the 7.7$\mu$m feature by \citet{Schweitzer06}.

Note that we do not fit Gaussian or Lorentzian profiles to measure PAH emissions \citep[e.g.,][]{Schweitzer06, Imanishi07} 
due to the relative weakness of PAH features in QSOs compared to their strong dust continuum. Since the three IR~QSOs 
in the QUEST sample (I~Zw~1, Mrk~1014, PG~1613+518) have 7.7/6.2 flux ratios of 4.2, 4.6, 4.9, respectively, similar to 
that of NGC~6240 (4.7, \citealt{Armus06}), we estimate the 6.2$\mu$m PAH fluxes for PG~QSOs in the \citet{Schweitzer06} 
sample (which has only 7.7$\mu$m measurements) by taking a 7.7/6.2 flux ratio of 4.7.

The fluxes of the ionised neon fine-structure lines in the MIR ([NeII], [NeV], [NeIII] at 12.81, 14.32, and 15.56$\mu$m) 
for IR~QSOs are measured based on the high-resolution IRS spectra, using the IDEA spectral analysis tool of the SMART 
software. The fluxes are measured by fitting a single Gaussian superposed on a local continuum approximated by a second-order 
polynomial. Flux upper limits (3$\sigma$) are derived adopting typical line widths of 600 \kms\, \citep{Schweitzer06}. 
The fluxes of the [NeII]12.81$\mu$m line for some ULIRGs in our sample are from \citet{Farrah07b}. 

\section{Results}

\begin{figure*}
\includegraphics[angle=90,scale=.8]{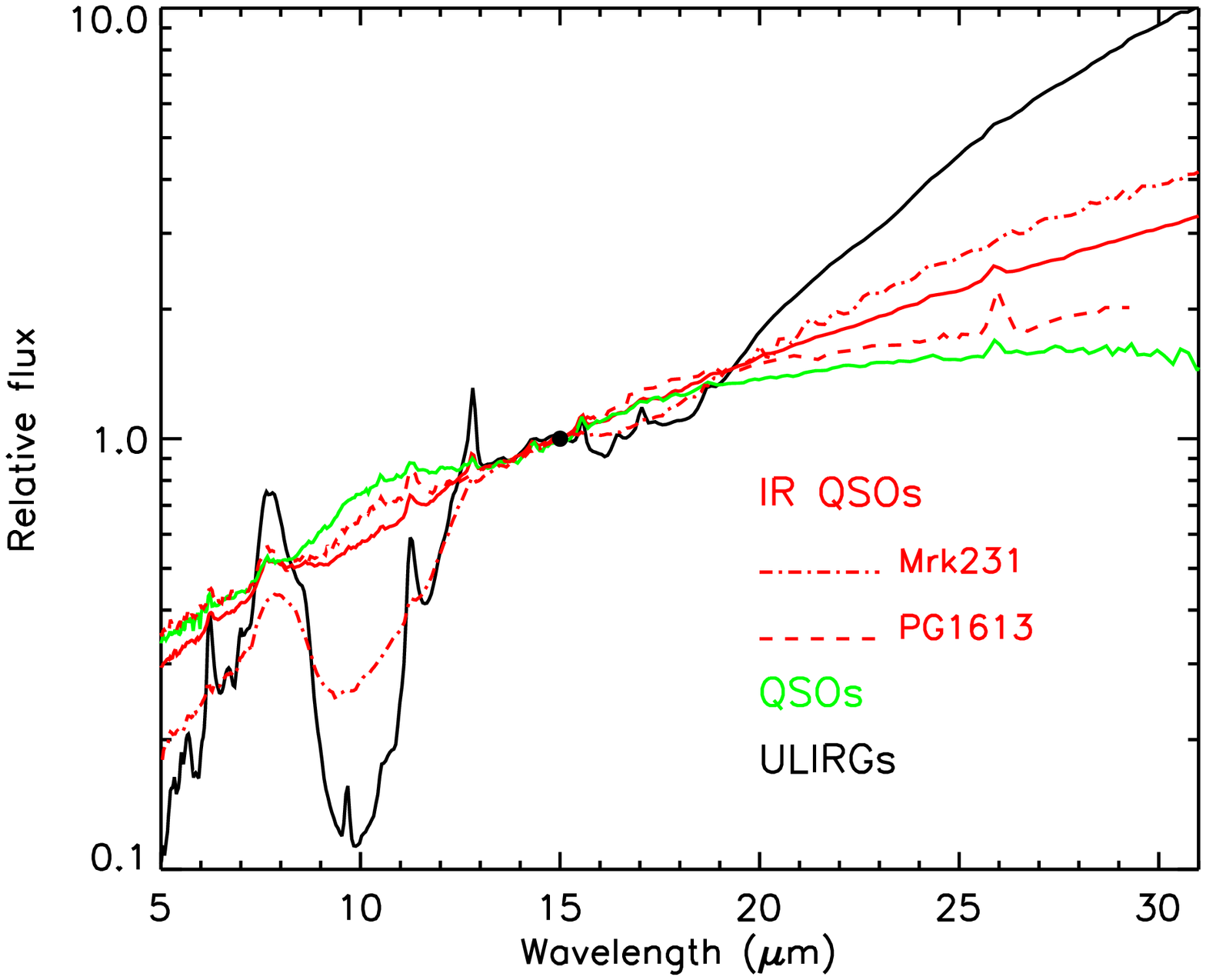}
\caption{Average MIR IRS low-resolution spectra of IR~QSOs (red solid line) in our sample, and QSOs (green), ULIRGs (black) 
from \citet{HL07}, normalised to f(15$\mu$m)=1 (denoted by a black dot). Spectra of two representative IR~QSOs (PG~1613+658 
and Mrk~231) are also shown by red dashed and red dot-dashed lines.} 
\label{fig_avgspec}
\end{figure*}

\subsection{Mid-infrared spectral characteristics of IR~QSOs} \label{sec:feature}

Fig.~\ref{fig_spec} shows the low-resolution {\it Spitzer} IRS MIR spectra of 18 IR~QSOs in our sample (except 
IRAS~F21219$-$1757 which has no LL mode observation). The dotted and dashed lines show the PAH features and 
the MIR fine-structure neon, sulphur, and oxygen lines. The shaded bar denotes the silicate emission/absorption 
feature centred at 9.7$\mu$m. One can see from these spectra that the PAH features at 6.2, 7.7, 8.6, 11.2 \& 12.7 $\mu$m 
and the MIR fine-structure emission lines, such as [NeII] 12.81$\mu$m, [NeV] 14.32$\mu$m, [NeIII] 15.56$\mu$m, 
[NeV] 24.32$\mu$m \& [OIV] 25.89$\mu$m are present in most IR~QSOs, although some emissions are weak for most of them. 
One can also see the silicate absorption feature at 9.7$\mu$m for several IR~QSOs (F00275, F13218, Mrk~231, F11119 
and F15462), which are rarely seen in PG~QSOs. For comparison, we show the average MIR spectra of IR~QSOs, PG~QSOs 
and ULIRGs in Fig.~\ref{fig_avgspec}. The average spectra of ULIRGs and PG~QSOs are from \citet{HL07}. It is clear 
from Fig.~\ref{fig_avgspec} that the slope of MIR continua from 15$\mu$m to 30$\mu$m of IR~QSOs is intermediate between 
that of ULIRGs and QSOs. We also show in the same figure the spectra of two representative IR~QSOs (PG~1613+658 and Mrk~231). 
Their MIR spectra are intermediate between the average spectra of ULIRGs and PG~QSOs. However, from the infrared, optical 
and X-ray observations, PG~1613+658 has the characteristics of classical QSOs \citep{Zheng02}, while Mrk~231 is an 
on-going merger with high SFR from its hundred-pc scale molecular disk \citep{Downes98}. 

In order to clarify more quantitatively the differences in the spectral properties, we study the properties of 6.2$\mu$m 
PAH, [NeII] 12.81$\mu$m luminosities, and MIR colour indices $\alpha$(30, 15), for IR~QSOs, PG~QSOs, and ULIRGs. The 
infrared colour index is defined as 
\begin{equation}
\alpha(\lambda_1, \lambda_2) = -{\log(F_{\nu}(\lambda_2)/F_{\nu}(\lambda_1)) \over \log(\lambda_2/\lambda_1)}.
\end{equation}
The central wavelengths (30 and 15$\mu$m) are selected in order to avoid the contamination by most strong spectral 
features (e.g., MIR fine-structure lines, PAH emissions, silicate emission/absorption features etc., see also 
\citealt{Brandl06, Schweitzer06}). 

\begin{figure*}
\includegraphics[angle=90,scale=.8]{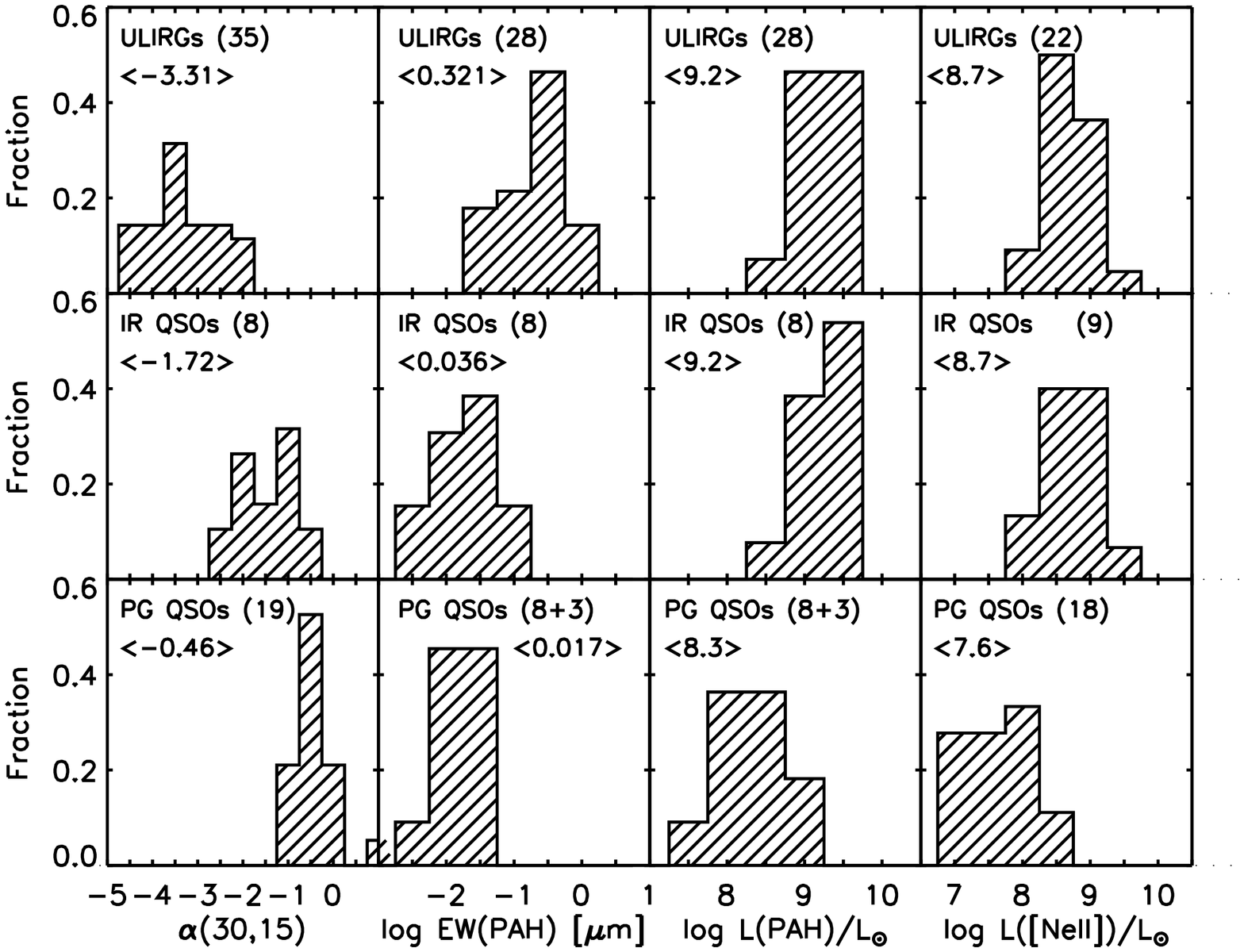}
\caption{From left to right: histograms of MIR colour index $\alpha$(30, 15), the EW (PAH 6.2$\mu$m), the 6.2$\mu$m PAH and 
12.8$\mu$m [NeII] line luminosities for ULIRGs (top), IR~QSOs (middle), and PG~QSOs (bottom) with detectable PAH and [NeII] 
emissions. The mean values are labelled in the panels, and the number of objects used in each histogram is in the bracket. 
Notice that the histogram of L([NeII]) for PG~QSOs is from the sample by \citet{Schweitzer06}, while the [NeII] luminosities 
for ULIRGs are derived from \citet{Farrah07b}.} 
\label{fig_hist}
\end{figure*}

\begin{figure*}
\includegraphics[angle=90,scale=.8]{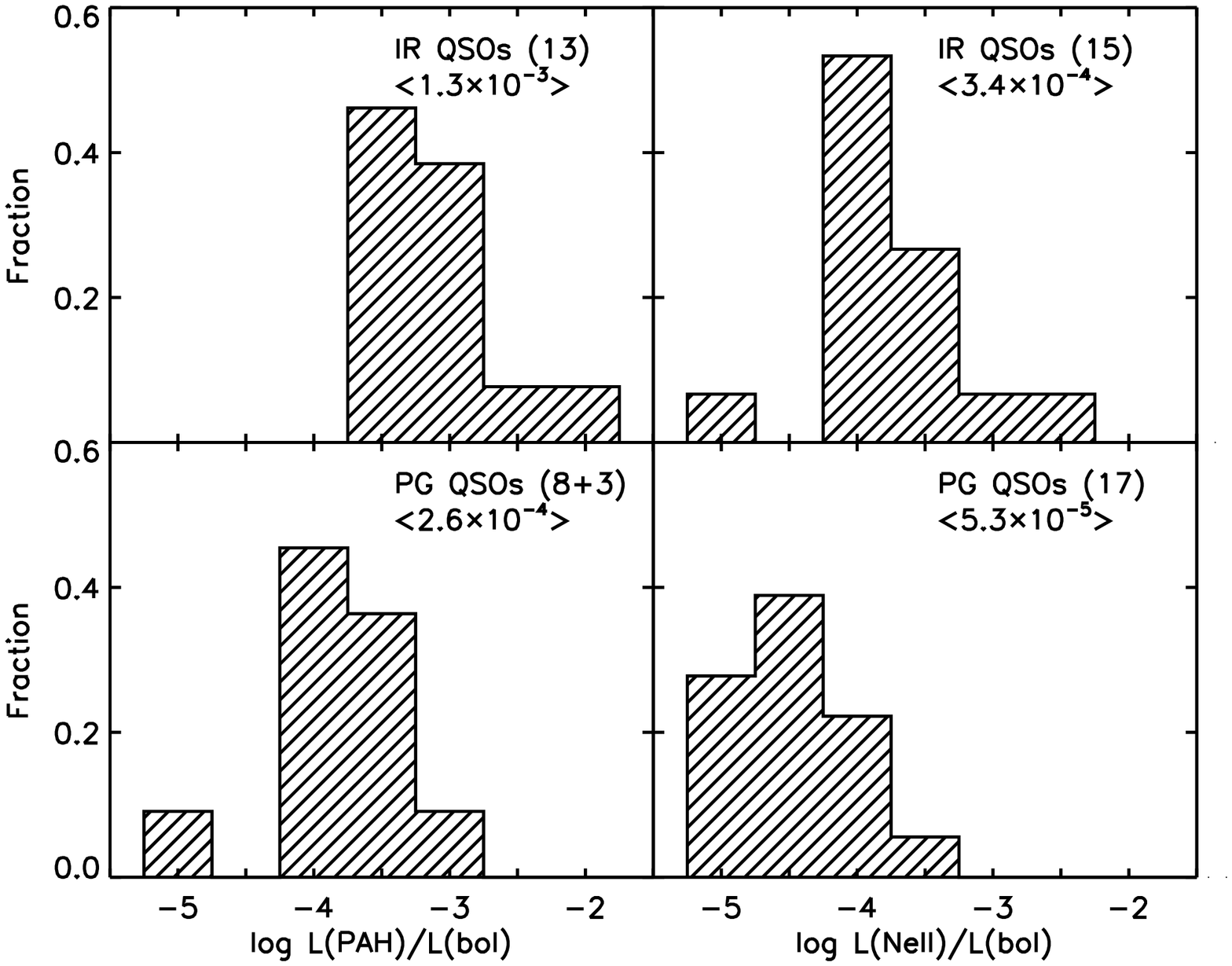}
\caption{Histograms of the 6.2$\mu$m PAH (left) and 12.8$\mu$m [NeII] line (right) luminosities normalised by the bolometric luminosity
for IR~QSOs (top) and PG~QSOs (bottom) with detectable PAH and [NeII] 
emissions. The mean values are labelled in the panels, and the number of objects used in each histogram is in the bracket. 
Notice that the histogram of L$_{\rm [NeII]}$/L$_{\rm bol}$ for PG~QSOs is from the sample of \citet{Schweitzer06}.}
\label{fig_hist2}
\end{figure*}

\begin{figure*}
\includegraphics[angle=90,scale=.75]{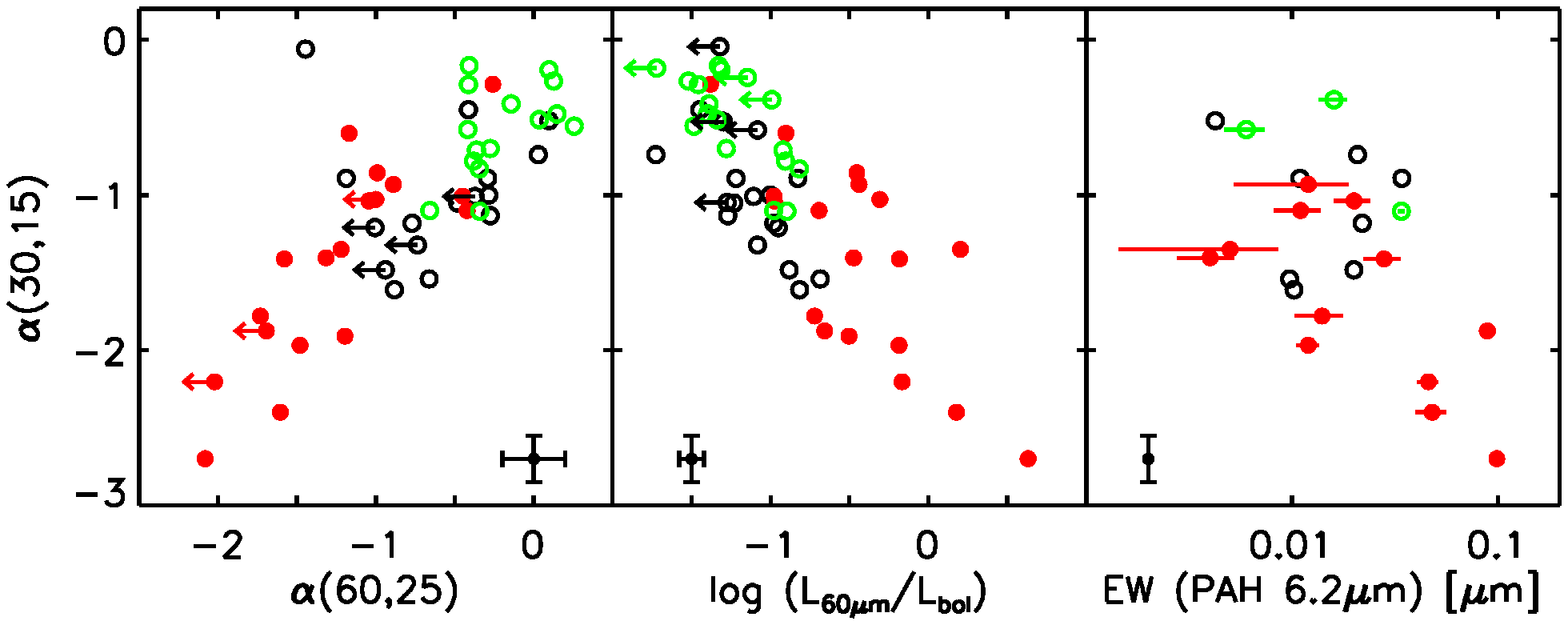}
\caption{From left to right: MIR colour index $\alpha$(30, 15) vs. infrared colour index $\alpha$(60, 25); vs. infrared 
excess (L$_{\rm 60\mu m}$/L$_{\rm bol}$); vs. EW (PAH 6.2$\mu$m) for IR~QSOs (red filled circle) and PG~QSOs (green and 
black open circle for PG~QSOs from our sample and \citealt{Schweitzer06}). The horizontal and/or vertical bars on the 
bottom right of each panel indicate the mean errors on the $\alpha$(60, 25), L$_{\rm 60\mu m}$/L$_{\rm bol}$, and 
$\alpha$(30, 15) values.} 
\label{fig_alpha}
\end{figure*}

The histograms for these quantities are shown in Fig.~\ref{fig_hist}. The mean values of $\alpha$(30, 15), EW (PAH 6.2$\mu$m), 
PAH 6.2$\mu$m and [NeII] 12.81$\mu$m luminosities are labelled at the top of each panel of Fig.~\ref{fig_hist}. From the 
left panels of Fig.~\ref{fig_hist}, the MIR spectral slopes for IR~QSOs are much flatter than those of ULIRGs, but are 
significantly steeper than those of PG~QSOs. Since classical QSOs have much lower FIR emissions than IR~QSOs at a given 
bolometric luminosity \citep{Hao05}, and $\alpha$(30, 15) is correlated with $\alpha$(60, 25) (see Fig.~\ref{fig_alpha} 
and \S\ref{sec:statis}), it is easy to understand why the $\alpha$(30, 15) slopes are steeper for IR~QSOs than those of 
classical QSOs (for details see \S\ref{sec:statis}). On the other hand, the emissions from hot dust heated by central AGN 
are mainly in the MIR band. Furthermore, {\it Spitzer} observations reveal that star formation regions in star-forming 
galaxies also contribute significantly to the MIR continuum emission \citep[e.g.,][]{Wu05, Calzetti07}. Therefore, for 
IR~QSOs, both starburst and central AGN contribute to the MIR continuum emission, leading to the MIR continuum of IR~QSOs 
being stronger than those of ULIRGs. Thus the slopes of MIR to FIR continuum of IR~QSOs are flatter than those of ULIRGs.

The histograms of EW (PAH 6.2$\mu$m) show that the mean value of EW (PAH 6.2$\mu$m) of IR~QSOs (0.031$\pm$0.024$\mu$m) 
is between those of ULIRGs (0.321$\pm$0.235$\mu$m) and PG~QSOs (0.017$\pm$0.008$\mu$m) and close to that of warm ULIRGs 
(0.04$\pm$0.05$\mu$m; \citealt{Desai07}). Considering the fraction of PAH detections for IR~QSOs and PG~QSOs being 70\% 
(13/19) and 40\% (11/27 in QUEST QSO sample) respectively, while the EW (PAH 6.2$\mu$m) for pure AGN is less than 
0.005-0.02$\mu$m \citep{Armus07}, the mean value of EW (PAH 6.2$\mu$m) of IR~QSOs is significantly larger than that 
of classical QSOs.
 
From the middle and right panels of Fig.~\ref{fig_hist} we see that the mean values of 6.2$\mu$m PAH and [NeII]12.81$\mu$m 
luminosities of IR~QSOs are similar to those of ULIRGs, but one order of magnitude higher than those of PG~QSOs. Apparently 
the PAH molecules in IR~QSOs can survive the FUV and X-ray photons emitted by central AGN, suggesting that they are shielded 
by a large amount of gas and dust from radiation \citep[see][]{Schweitzer06}. 

Since the bolometric luminosities (L$_{\rm bol}$) of PG~QSOs observed by IRS of {\it Spitzer} are systematically smaller than 
those of IR~QSOs, where L$_{\rm bol}$ is calculated from the optical continuum emission ($L_{\rm bol} \approx 9 \lambda 
L_{\rm \lambda}(5100\AA)$, \citealt{Hao05}), the differences in their IR continuum shapes may simply arise because more optically 
luminous QSOs have more extended dust tori. To check this possibility, we investigated the 6.2$\mu$m PAH and [NeII]12.81$\mu$m 
luminosities normalised by L$_{\rm bol}$. Fig.~\ref{fig_hist2} shows the distributions of L$_{\rm [PAH]}$/L$_{\rm bol}$ and 
L$_{\rm [NeII]}$/L$_{\rm bol}$ for IR~QSOs and PG~QSOs, respectively. It is clear from Fig.~\ref{fig_hist2} that the mean values 
of L$_{\rm [PAH]}$/L$_{\rm bol}$ and L$_{\rm [NeII]}$/L$_{\rm bol}$ for IR~QSOs are also about one order of magnitude higher than 
those of PG~QSOs. Therefore, the different properties of 6.2$\mu$m PAH and [NeII]12.81$\mu$ emissions between IR~QSOs and PG~QSOs 
are unlikely from different dust tori, instead the differences arise because of different star formation properties (see section 4.2).

In summary, the mid-IR spectroscopic properties, including the continuum slope and emission line strengths, of IR~QSOs, PG~QSOs 
and ULIRGs are consistent with that IR~QSOs are in a transitional phase from ULIRGs to classical QSOs, confirming the results 
from previous studies \citep{CS01, Hao05}.

\subsection{Statistics on spectral parameters} \label{sec:statis}

In this subsection, we will use the MIR spectroscopic features, including the MIR continuum slope $\alpha$(30, 15), 6.2$\mu$m 
PAH and fine-structure emission lines, to investigate the origin of MIR emissions of IR~QSOs. We will also use these 
properties, combined with FIR and optical properties, to disentangle the starburst and AGN contributions in these objects.

Fig.~\ref{fig_alpha} shows the relations of $\alpha$(30, 15) vs. $\alpha$(60, 25) (left panel), $\alpha$(30, 15) vs. the FIR 
excess L$_{\rm 60\mu m}$/L$_{\rm bol}$ (middle panel), and $\alpha$(30, 15) vs. the EW (PAH 6.2$\mu$m) (right panel) for both 
IR~QSOs and PG~QSOs. 
It is clear from Fig.~\ref{fig_alpha} that the 
colour indices of $\alpha$(30, 15) and $\alpha$(60, 25) are closely correlated, indicating that $\alpha$(30, 15) can express 
the relative strength of FIR to MIR emission for QSOs. The middle and right panels of Fig.~\ref{fig_alpha} show the 
correlations between $\alpha$(30, 15) with FIR excess L$_{\rm 60\mu m}$/L$_{\rm bol}$, and between $\alpha$(30, 15) with the 
EW (PAH 6.2$\mu$m). The EW (PAH 6.2$\mu$m) is the ratio of PAH 6.2$\mu$m emission line to $\sim$6$\mu$m continuum. Since 
the 6.2$\mu$m PAH emission is from star formation regions, and the 6$\mu$m continuum traces the AGN contribution 
\citep[e.g.,][]{Gallagher07}, thus the EW (PAH 6.2$\mu$m) expresses the relative contribution of star formation to AGN 
\citep{Schweitzer06, Armus07}. In fact, \citet{Desai07} also found the strong correlation between infrared spectral slope 
and the EW (PAH 6.2$\mu$m) for ULIRGs, especially for ULIRGs with Seyfert~1 and Seyfert~2 optical spectra, while our 
results extend such relation to infrared luminous QSOs and PAH detected PG~QSOs. We conclude that $\alpha$(30, 15), FIR 
excess and EW (PAH 6.2$\mu$m) can serve as indicators of the relative contributions of starbursts to AGNs (\citealt{Hao05} 
and see below). 

\citet{Ho07} suggest that the ionised neon fine-structure lines [NeII]12.81$\mu$m and [NeIII]15.56$\mu$m can be used as 
a SFR indicator for star-forming galaxies. \citet{Farrah07b} extend this relation to ULIRGs. In addition, \citet{Schweitzer06} 
found a strong correlation between the far-infrared continuum (L$_{\rm 60\mu m}$) and low-ionisation [NeII] line emission 
for both PG~QSOs and ULIRGs, and argued that the [NeII] line can also be used to estimate the SFR in QSO host galaxies. 
One advantage to use [NeII] 12.81$\mu$m as a SFR estimator is that it suffers much less extinction than optical lines, 
such as $H\alpha$ and [OII]3727\AA. However, there is still a debate about the origin of [NeII] emission, because the narrow 
line region of QSOs may also contribute substantially \citep[e.g.,][]{Ho07}. Therefore, it is worth investigating the origin 
of [NeII]12.81$\mu$m emission for IR~QSOs by comparing the multi-wavelength properties of PG~QSOs, IR~QSOs and ULIRGs.

\begin{figure*}
\includegraphics[angle=90,scale=.8]{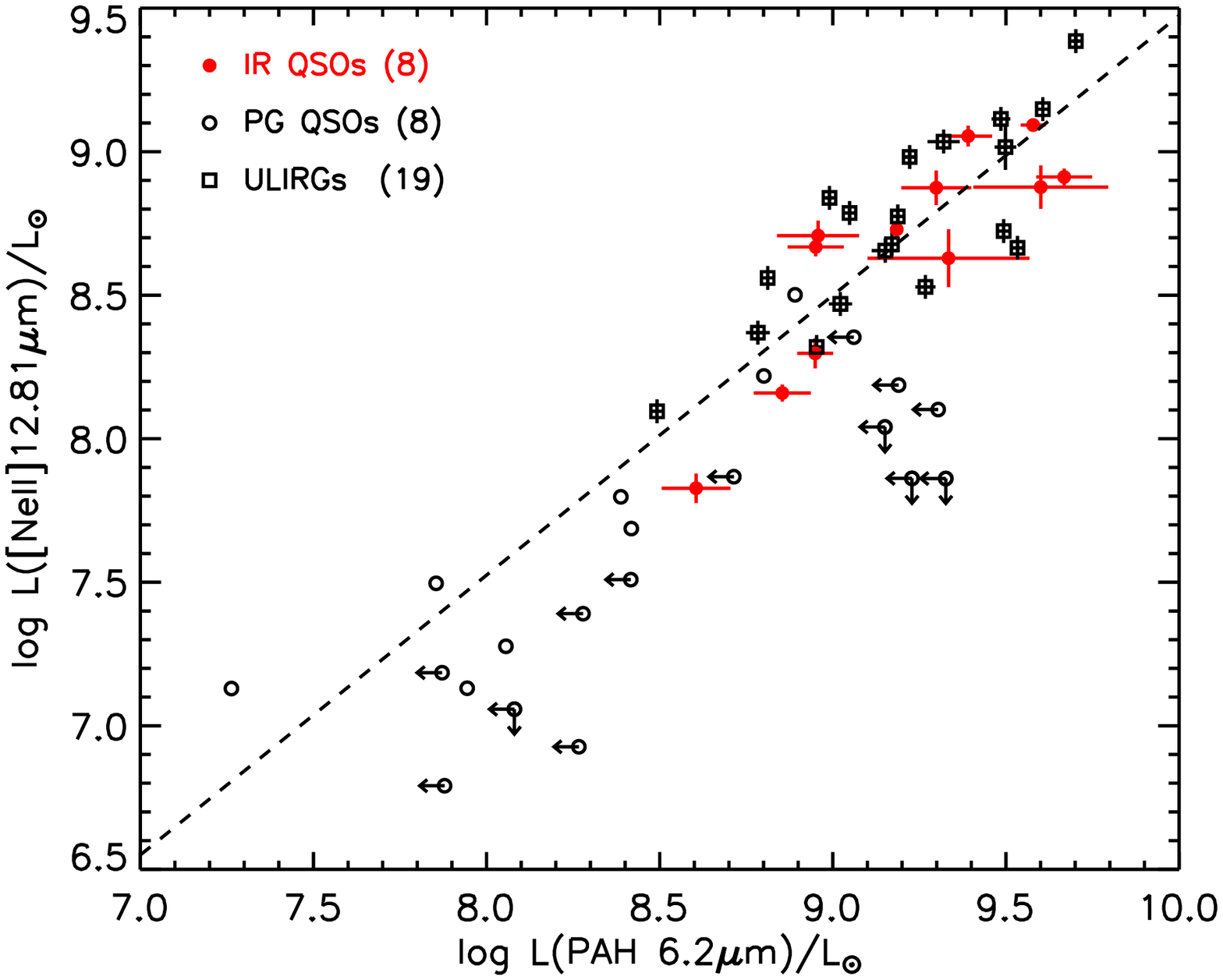}
\caption{The PAH 6.2$\mu$m vs. [NeII] 12.81$\mu$m luminosities for IR~QSOs, PG~QSOs (from \citealt{Schweitzer06}), and ULIRGs 
in our sample which have [NeII] measurements by \citet{Farrah07b}. The number of each type of galaxies is indicated in the 
bracket, while the dashed line represents the linear regression for all sample objects. Note we excluded objects with undetected 
(upper-limits) PAH or [NeII] emissions in the regression analysis. The Spearman Rank-order (S-R) correlation analysis gives 
the coefficient of linear regression as 0.89 with significance of $>$ 99.99$\%$ for the above correlation.} 
\label{fig_ne2pah}
\end{figure*}

Fig.~\ref{fig_ne2pah} shows the relation between PAH 6.2$\mu$m and [NeII]12.81$\mu$m luminosities for IR~QSOs, ULIRGs and 
PG~QSOs with firmly detected [NeII]12.81$\mu$m and PAH 6.2$\mu$m emissions. Because the PAH emissions are purely from star 
formation regions \citep{Shi07}, the tight correlation between 6.2$\mu$m PAH and [NeII]12.81$\mu$m luminosities (at a 
statistical level of $>$ 99.99$\%$ with the Spearman Rank-order test) demonstrates that (at least part of) the [NeII] 
12.81$\mu$m emission is also from star formation regions. Note that for the PG~QSOs shown in Fig.~\ref{fig_ne2pah}, their 
[NeII] 12.81$\mu$m luminosities normalised by the bolometric luminosities of AGNs (L$_{\rm [NeII]}$/L$_{\rm bol}$ ratios, 
see below) are about three times higher than that of PAH undetected PG~QSOs (see also \citealt{Schweitzer06}). Therefore, 
it is likely that the star formation contributes significantly to the [NeII] emission not only for IR~QSOs, but also 
for PG~QSOs with detectable PAH emissions \citep{Netzer07}.

The mean values of L$_{\rm [NeII]12.81\mu m}$/L$_{\rm bol}$ ratios are $3.4\pm3.5 \times 10^{-4}$, $5.3\pm3.6 \times 10^{-5}$, 
$8.1\pm5.2 \times 10^{-5}$ and $2.7\pm1.4 \times 10^{-5}$ for IR~QSOs, PG~QSOs, PAH-detected and PAH-undetected PG~QSOs, 
respectively. Thus for the same bolometric luminosity of a central AGN, the mean [NeII]12.81$\mu$m luminosity of IR~QSOs 
is about one order of magnitude higher than that of classical QSOs. Taken together with the tight correlation between 
[NeII] and PAH luminosities (see above), we conclude that the [NeII]12.81$\mu$m emission of IR~QSOs is mainly from star 
formation, while the contribution from the narrow line region of AGNs is not significant ($\lesssim 10\%$).

\section{Discussion}
By comparing the MIR spectroscopic properties of IR~QSOs, ULIRGs and PG~QSOs, we found that the indicators of relative 
contributions of starbursts to AGNs (such as colour index $\alpha$(30, 15) and EW [PAH 6.2$\mu$m]) for IR QSOs are between 
those of ULIRGs and PG QSOs. These results are consistent with the findings of \citet{CS01} and \citet{Hao05} that (at 
least some) infrared luminous QSOs (IR~QSOs) are at a transitional stage from ULIRGs to classical QSOs. Below we consider 
the star formation rates and AGN/star formation feedback in more detail.

\subsection{SFR determined by the [NeII]12.81$\mu$m and PAH luminosities}

As we argued, the AGN contribution to [NeII] emission for ULIRGs and IR~QSOs is probably very small, and there is a tight 
correlation between [NeII]12.81$\mu$m and PAH 6.2$\mu$m luminosities (see \S\ref{sec:statis}). We examine in more detail 
how they can be used as approximate SFR indicators for ULIRGs and IR~QSOs.

\begin{figure*}
\includegraphics[angle=0,scale=.7]{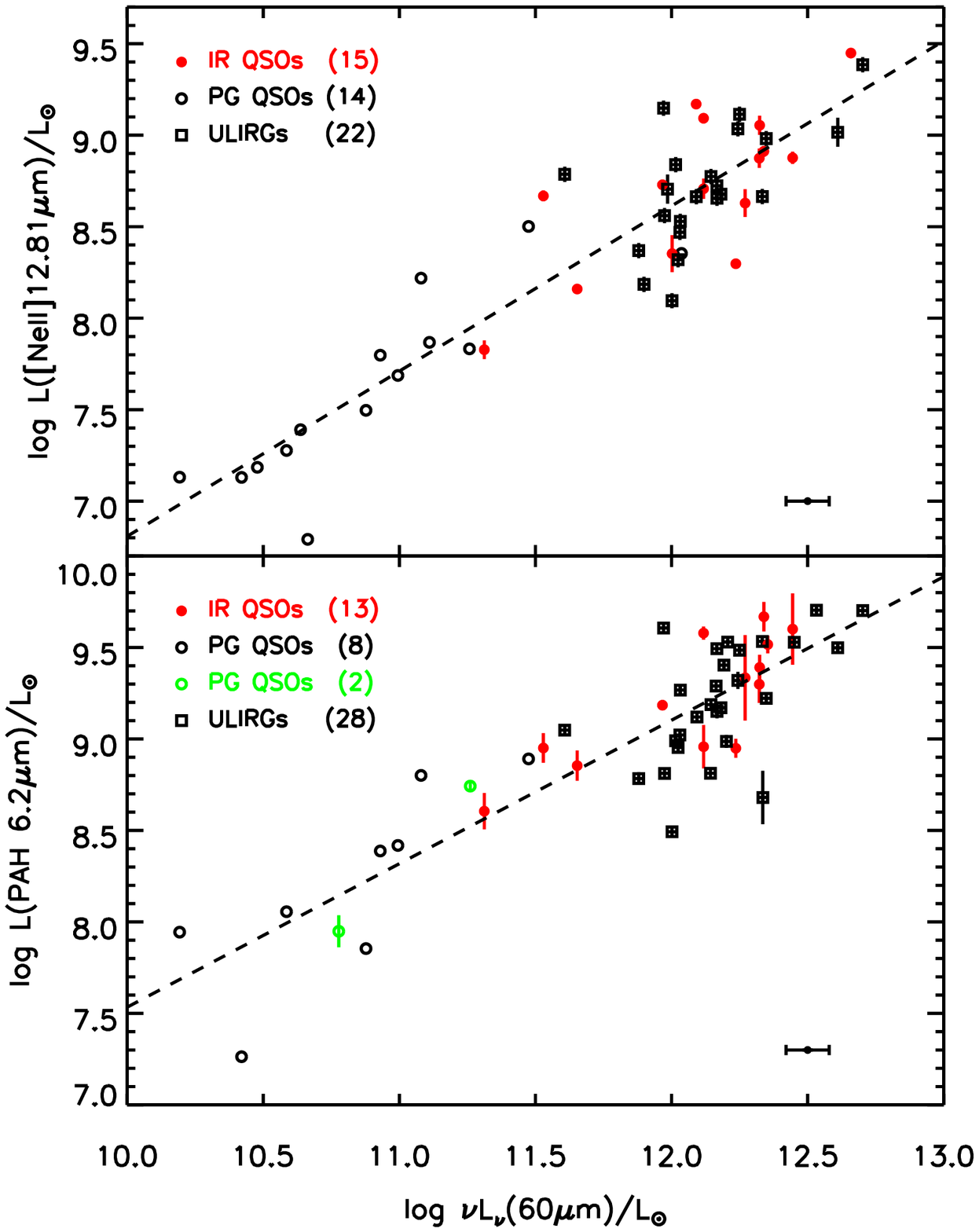}
\caption{Top panel: Luminosities L(60$\mu$m) vs. L([NeII]12.81$\mu$m) for IR~QSOs, PG~QSOs (from \citealt{Schweitzer06}), and 
ULIRGs with detectable [NeII] and 60$\mu$m emissions. Bottom panel: Luminosities L(60$\mu$m) vs. L(PAH 6.2$\mu$m) for IR~QSOs, 
PG~QSOs in our sample (green) and from \citet{Schweitzer06} (black), and ULIRGs with detectable PAH and 60$\mu$m emissions. The 
6.2$\mu$m PAH fluxes for PG~QSOs from \citet{Schweitzer06} were estimated by assuming a 7.7/6.2 flux ratio of 4.7 derived from 
the measurement of the ULIRG NGC~6240 with a buried AGN \citep{Armus06}. The number of each galaxy type is indicated in the 
bracket, while the dashed lines represent the linear regressions for the above two correlations. The horizontal bar on the bottom 
right of each panel indicates the mean error of the 60$\mu$m luminosity.} 
\label{fig_sfr}
\end{figure*}

Fig.~\ref{fig_sfr} shows the [NeII]12.81$\mu$m (top panel) and PAH 6.2$\mu$m (bottom panel) luminosities versus the 60$\mu$m 
luminosity for IR~QSOs, PG~QSOs and ULIRGs with firmly detected [NeII]12.81$\mu$m and PAH 6.2$\mu$m emissions. A Spearman 
Rank-order analysis show that both correlations are significant at $>$ 99.99$\%$ level. The dashed lines in Fig.~\ref{fig_sfr} 
represent the least-squares regression fits:
\begin{equation}
\log L_{\rm NeII} = (0.90\pm0.06)\log L_{\rm 60\mu m} - (2.22\pm0.69),
\end{equation}
\begin{equation}
\log L_{\rm PAH} = (0.78\pm0.06)\log L_{\rm 60\mu m} - (0.31\pm0.74).
\end{equation}
The fitting formula (2) is consistent with that of \citet{Ho07} for star-forming galaxies (within the large errors). Note 
that our sample objects have much higher $60\mu{\rm m}$ and [NeII] luminosities than their star-forming galaxies. Thus both 
[NeII] 12.81$\mu$m and PAH 6.2$\mu$m luminosities can be used as approximate SFR indicators not only for normal star-forming 
galaxies, but also for galaxies with high infrared luminosities, such as ULIRGs and IR~QSOs (see \citealt{Brandl06} and 
\citealt{Farrah07b}).

However, the mean scatters (about 0.7 to 0.8 dex) in the relation of [NeII]12.81$\mu$m, PAH 6.2$\mu$m with $60\mu{\rm m}$ 
luminosities are larger than that (about 0.6 dex) of star-forming galaxies with lower infrared luminosity (\citealt{Ho07}). 
Comparing Fig.~\ref{fig_sfr} with Fig.~\ref{fig_ne2pah}, one can see that the scatter in the relation of [NeII]12.81$\mu$m 
vs. PAH 6.2$\mu$m luminosities (about 0.6 dex) is smaller than that in the relations of [NeII]12.81$\mu$m, PAH 6.2$\mu$m 
luminosities with L(60$\mu$m). It is also clear that most large scatters are from ULIRGs. This is perhaps not surprising 
since the range in the 9.7$\mu$m silicate absorption depth among ULIRGs is quite large (see \citealt{Armus07}, \citealt{Spoon07}). 
In short, the large scatters for ULIRGs seen in the relations may be due to complicated, patchy extinctions among these 
galaxies in the MIR band. A detailed discussion on extinction for ULIRGs can be found in \citet{Farrah07b}.
                                           
\subsection{AGN/Star formation feedback in the transitional stage}

\begin{figure*}
\includegraphics[angle=90,scale=.8]{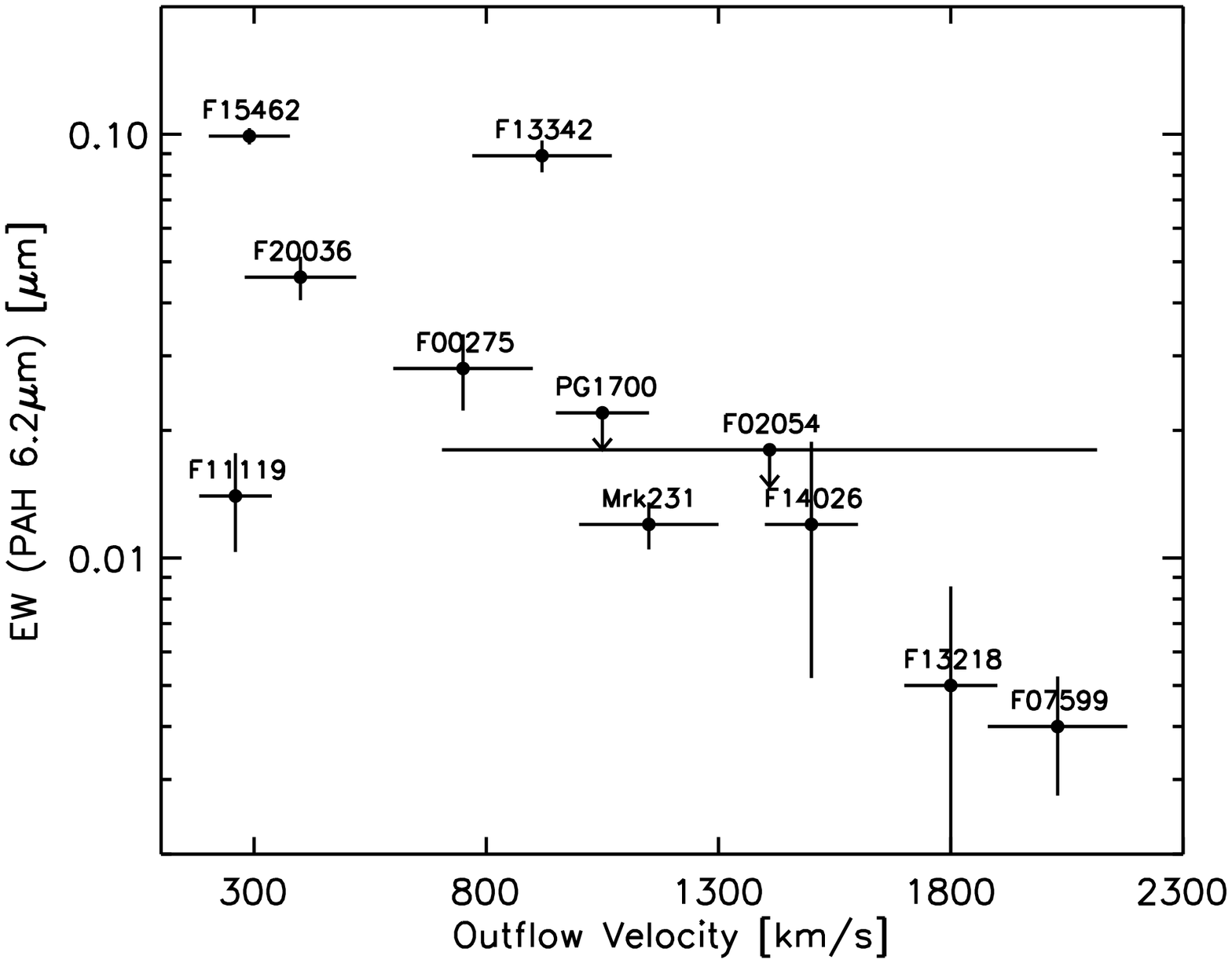}
\caption{The velocity of outflows vs. the equivalent width of the 6.2$\mu$m PAH emission feature (EW (PAH 6.2$\mu$m)) for 
eleven IR~QSOs (black filled circle) which show obvious H$\beta$ blueshifts. The Kendalls ($\tau$) rank correlation analysis 
gives coefficient of linear regression as $-0.66$ with a significance of 99.5$\%$. The outflow velocities were derived from 
the H$\beta$ emission line profiles \citep{Zheng02, Lipari05}.} 
\label{fig_outflow}
\end{figure*}

One explanation for the observed correlation between spheroidal and black hole mass \citep[e.g.,][]{Magorrian98, Ferrarese05} 
is that star formation and the growth of central black holes may be self-regulated: AGNs/star formation can drive nuclear 
outflows which in turn suppress further cooling and star formation (\citealt{Silk98}). While the detailed processes are still 
to be understood, it is now increasingly clear that feedback and outflows play an important role in galaxy formation and evolution.

So far most observational evidence for AGN feedback is from radio observations at the centre of clusters or groups of galaxies 
\citep{Batcheldor07}. On the galaxy scale, evidence is still limited. As discussed above, IR~QSOs have high SFRs and accretion 
rates \citep{Hao05}, outflow properties in these objects may thus provide hints on the feedback processes on galaxy or group scale. 

It is well known that low-ionisation broad absorption line QSOs (loBAL~QSOs) comprise about 15\% of BAL~QSO population. They 
are defined as a subclass of BAL~QSOs with an obvious blueshifted absorption in Mg II $\lambda$$\lambda$2795,2802 and other 
low-ionisation species \citep{Weymann91}. Such absorption troughs arise from resonance absorption by outflowing gas and dust 
\citep{Voit93}. In addition, there is a rare class of loBAL~QSOs, showing absorption features from excited iron (termed as 
FeLoBAL~QSOs). The outflow velocities for most BAL~QSOs span a large range, up to a few times $10^{4}$ \kms, which may be formed 
on a scale of $<1\,{\rm pc}$ and directly associated with the wind from an accretion disk or molecular torus \citep{Weymann85}. 
However, recent spectral analyses based on Keck observations for LoBAL~QSOs or FeLoBAL~QSOs reveal that the outflow velocities 
of some LoBAL~QSOs range from several hundred to several thousand \,\kms\, and the wind is from regions of a few hundred pc 
\citep[e.g.,][]{Ganguly08, Hamann00, deKool02}, which is much larger than the central engine of AGNs, but similar to the size 
of the nuclear starburst region of ULIRGs \citep{Downes98}. 
 
On the other hand, \citet{CS02} studied four loBAL~QSOs at z$<$0.4 (Mrk~231, IRAS~14026$+$4341, IRAS~F07599+6508, PG~1700+518; 
all four are in our sample) and found that all are ULIRGs with merging signatures. They argued that loBAL~QSOs cannot simply 
be explained by orientation effects, rather, they are directly related to young systems, still surrounded by gas and dust. 
It represents a short stage in the early life of a large fraction of QSOs. Moreover, \citet{Farrah07a} reported the detections 
of mid/far-infrared emission from 9 FeLoBAL QSOs by MIPS. They found that all of their objects are IR bright with infrared 
luminosities as high as ULIRGs. Thus all loBAL~QSOs and FeLoBAL QSOs with MIR to FIR information are infrared ultra-luminous, 
suggesting a link among loBAL~QSOs, FeLoBAL QSOs and IR~QSOs.

For loBAL~QSOs or FeLoBAL QSOs, the velocities of outflows can be measured from blueshifted absorption lines. Unfortunately it 
is still unclear whether most IR~QSOs are loBAL~QSOs or FeLoBAL QSOs, because not many IR QSOs have been observed in the UV. 
However, there are measurements for the blueshifts in the permitted optical emission lines for our IR~QSO sample \citep{Zheng02}. 
The line profiles often contain two Gaussian components, one broad and one narrow. The outflow velocities are determined by the 
blueshift of the broad Gaussian component relative to the narrow component of permitted emission lines (H$\beta$) for IR~QSOs 
\citep[see][]{Zheng02}. We assume the blueshifted broad Gaussian component is related to the outflow of clouds 
\citep{Leighly01, Batcheldor07}. 

Fig.~\ref{fig_outflow} shows the correlation between EW (PAH 6.2$\mu$m) and outflow velocities for IR~QSOs. As the outflow 
velocity increases, the EW (PAH 6.2$\mu$m) decreases. Since the EW (PAH 6.2$\mu$m) expresses the relative contribution of 
star formation to AGN in QSOs (see $\S$4.2), the correlation shown in Fig.~\ref{fig_outflow} implies that QSOs with higher 
outflow velocities have a lower ratio of SFR to accretion rate. It may be because more energetic AGNs and central massive 
starbursts can drive faster outflows which then suppress or even quench star formation by heating up or expelling the cold 
gas and dust in the QSO hosts, leading to a lower ratio of SFR to accretion rate. Another possibility is that the higher 
velocity outflows are more efficient in ejecting dust cocoons of AGNs, leading to AGNs becoming the dominant source compared 
with starbursts. We should caution, however, that the sample shown in Fig.~\ref{fig_outflow} is small; a larger sample would 
allow us to better understand the relative importance of star formation and AGN in feedback and driving outflows.

\section{Summary}
In this work, we studied the MIR spectral properties of low-redshift IR~QSOs based on spectroscopic observations with 
{\it Spitzer} IRS, and compared their properties with those of ULIRGs and optically-selected PG~QSOs. The following are 
our main results:

1. The average MIR spectra, MIR spectral slopes, 6.2$\mu$m PAH emission strengths and [NeII] 12.81$\mu$m luminosities 
of IR~QSOs are distinct from those of PG~QSOs. The MIR properties of IR~QSOs are intermediate between ULIRGs and 
optically-selected QSOs, indicating that IR~QSOs are at a transitional stage from ULIRGs to classical QSOs.

2. From the correlations between $\alpha$(30, 15) with $\alpha$(60, 25), FIR excess L$_{\rm 60\mu m}$/L$_{\rm bol}$, and 
EW (PAH 6.2$\mu$m) for both IR~QSOs and PG~QSOs, we find that the MIR colour index $\alpha$(30, 15) is a good indicator 
of the relative contribution of starbursts to AGNs for all QSOs.

3. Both PAH and [NeII]12.81$\mu$m luminosities of IR~QSOs are in the same range as those of ULIRGs, but are one order of 
magnitude higher than those of PG~QSOs (see Fig.~\ref{fig_hist}). From the tight correlation between PAH 6.2$\mu$m and [NeII] 
12.81$\mu$m luminosities for IR~QSOs, ULIRGs and PAH detected PG~QSOs, and the fact that the L$_{\rm [NeII]}$/L$_{\rm bol}$ 
ratio for IR~QSOs is about one order of magnitude higher than that of classical QSOs, we conclude that the [NeII] 12.81$\mu$m 
emission of IR~QSOs is dominated by star formation, thus their luminosity can be used as an approximate indicator of their SFRs.

4. Outflows in IR~QSOs play an important role in suppressing star formation by heating and/or expelling cold gas surrounding 
nuclei of QSOs. Thus IR~QSOs are an important observational sample to investigate AGN and star formation feedback processes.

\section*{Acknowledgements}
We thank the anonymous referee for constructive comments and suggestions. We acknowledge Drs. J. Wang, D.-W. Xu, 
Y. Shi, S. Veilleux for advice and helpful discussions. We also thank Z.-H. Shang, Y.-L. Wu, K. Zhang for helpful 
discussion on the {\it Spitzer} IRS data reductions, and Sarah Bryan for a careful reading of the manuscript. 
CC acknowledges the support of the Tianjin Astrophysics Center (TAC) at Tianjin Normal University, and the full 
living and travel supports of the European Union Marie Curie Program for him to participate in the AGN/VLTI 
summer school held in Poland, 2007. SM acknowledges the travel support of the NSFC, the Chinese Academy of Sciences 
and the Tianjin municipal government. This work is based on observations made with the {\it Spitzer Space Telescope}, 
which is operated by the Jet Propulsion Laboratory, California Institute of Technology under a contract with NASA. 
The IRS was a collaborative venture between Cornell University and Ball Aerospace Corporation funded by NASA through 
the Jet Propulsion Laboratory and Ames Research Center. SMART was developed by the IRS Team at Cornell University 
and is available through the Spitzer Science Center at Caltech. This research has made use of the NASA/IPAC 
Extragalactic Database (NED) which is operated by the Jet Propulsion Laboratory, California Institute of Technology, 
under contract with the National Aeronautics and Space Administration. This project is supported by the NSF of China 
10333060, 10778622 and 973 Program No.2007CB815405, 2007CB815406.


\clearpage

%
\begin{table*}
 \centering
 \begin{minipage}{140mm}
  \caption{QSO Sample}
  \begin{tabular}{@{}lcccccc@{}}
  \hline
~ & R.A. & Decl. & ~ & ~ & ~ & ~ \\
Object & (J2000.0) & (J2000.0) & Redshift & log$\left (L_{\rm 60\mu{\rm m}} / L_\odot \right)$ 
& log$\left (L_{\rm opt} / L_\odot \right)$ & SpecType\\
~~(1) & (2) & (3) & (4) & (5) & (6) & (7)\\
  \hline
\multicolumn{7}{c}{IR~QSOs}\\
  \hline
I~Zw~1 & 00~53~34.9 & $+$12~41~36 & 0.0611 & 11.310 & 11.050 & S1\\
F00275$-$2859 & 00~30~04.2 & $-$28~42~25 & 0.2781 & 12.342 & 11.568 & QSO\\
3C~48 & 01~37~41.3 & $+$33~09~35 & 0.3670 & 12.648 & 11.991 & QSO\\
Mrk~1014 & 01~59~50.2 & $+$00~23~41 & 0.1630 & 12.326 & 11.192 & QSO\\
F02054+0835 & 02~08~06.8 & $+$08~50~02 & 0.3450 & 12.466 & 11.819 & QSO\\
F07599+6508 & 08~04~33.1 & $+$64~59~49 & 0.1483 & 12.116 & 11.637 & QSO\\
F11119+3257 & 11~14~38.9 & $+$32~41~33 & 0.1890 & 12.322 & 12.089 & QSO\\
3C~273 & 12~29~06.7 & $+$02~03~09 & 0.1583 & 12.263 & 12.427 & QSO\\
Mrk~231 & 12~56~14.2 & $+$56~52~25 & 0.0422 & 12.236 & 11.467 & QSO\\
F13218+0552 & 13~24~19.9 & $+$05~37~05 & 0.2051 & 12.270 & 11.467 & QSO\\
F13342+3932 & 13~36~24.1 & $+$39~17~31 & 0.1793 & 12.116 & 11.821 & QSO\\
F14026+4341 & 14~04~38.8 & $+$43~27~07 & 0.3233 & 12.445 & 11.930 & QSO\\
PG~1543+489 & 15~45~30.2 & $+$48~46~09 & 0.3996 & 12.344 & 11.843 & QSO\\
F15462$-$0450 & 15~48~56.8 & $-$04~59~34 & 0.0998 & 11.995 & 10.381 & S1\\
PG~1613+658 & 16~13~57.2 & $+$65~43~10 & 0.1290 & 11.533 & 11.550 & QSO\\
PG~1700+518 & 17~01~24.8 & $+$51~49~20 & 0.2920 & 12.090 & 12.115 & QSO\\
F18216+6419 & 18~21~57.3 & $+$64~20~36 & 0.2970 & 12.659 & 12.607 & QSO\\
F20036$-$1547 & 20~06~31.7 & $-$15~39~08 & 0.1919 & 12.359 & 11.566 & QSO\\
F21219$-$1757 & 21~24~41.6 & $-$17~44~46 & 0.1120 & 11.661 & 10.952 & S1\\
  \hline
\multicolumn{7}{c}{PG~QSOs}\\
  \hline
PG~0804+761 & 08~10~58.6 & $+$76~02~42 & 0.1000 & 10.794 & 11.360 & QSO\\
PG~0838+770 & 08~44~45.2 & $+$76~53~09 & 0.1310 & 10.994 & 11.020 & S1\\
PG~0844+349 & 08~47~42.4 & $+$34~45~04 & 0.0640 & 10.323 & 10.854 & S1\\
PG~1004+130 & 10~07~26.1 & $+$12~48~56 & 0.2408 & 11.639 & 11.964 & QSO\\
PG~1119+120 & 11~21~47.1 & $+$11~44~18 & 0.0502 & 10.536 & 10.509 & S1\\
PG~1149$-$110 & 11~52~03.5 & $-$11~22~24 & 0.0490 & 10.435 & 10.299 & S1\\
PG~1211+143 & 12~14~17.7 & $+$14~03~13 & 0.0809 & 11.084 & 11.414 & QSO\\
PG~1351+640 & 13~53~15.8 & $+$63~45~45 & 0.0882 & 11.248 & 11.203 & QSO\\
PG~1411+442 & 14~13~48.3 & $+$44~00~14 & 0.0896 & 10.579 & 10.945 & S1\\
PG~1426+015 & 14~29~06.6 & $+$01~17~06 & 0.0865 & 10.753 & 11.194 & QSO\\
PG~1444+407 & 14~46~45.9 & $+$40~35~06 & 0.2673 & 11.556 & 11.636 & QSO\\
PG~1501+106 & 15~04~01.2 & $+$10~26~16 & 0.0364 & 10.468 & 10.649 & S1\\
PG~1519+226 & 15~21~14.2 & $+$22~27~43 & 0.1370 & 10.963$^{a}$ & 11.001 & S1\\
PG~1534+580 & 15~35~52.3 & $+$57~54~09 & 0.0296 & 9.663 & 10.069 & S1\\
PG~1535+547 & 15~36~38.3 & $+$54~33~33 & 0.0389 & 9.742$^{a}$ & 10.530 & S1\\
PG~1552+085 & 15~54~44.6 & $+$08~22~22 & 0.1190 & 10.782$^{a}$ & 10.974 & S1\\
PG~1612+261 & 16~14~13.2 & $+$26~04~16 & 0.1309 & 11.053 & 11.019 & S1\\
PG~1617+175 & 16~20~11.3 & $+$17~24~28 & 0.1124 & 10.637 & 11.140 & QSO\\
PG~2130+099 & 21~32~27.8 & $+$10~08~19 & 0.0630 & 10.748 & 10.876 & S1\\
\hline
\end{tabular}
~\\
~\\
Notes -- Basic properties of IR~QSOs and PG~QSOs in our sample. Col.(1): source name; Col.(2)-(3): right ascension 
(hours, minutes, seconds) and declination (degrees, arcminutes, arcseconds), from NED; Col.(4): redshift, from NED; 
Col.(5): monochromatic luminosity at 60$\mu$m; Col.(6): monochromatic luminosity at 5100\AA, $\lambda$L$_{\lambda}$(5100\AA). 
The values are derived from \citet{Hao05} except for F14026$+$4341 it is measured based on the SDSS spectrum, and for 
some PG~QSOs not included in \citet{Hao05} their values were taken from \citet{Neugebauer87} and calculated in the same 
manner as in \citet{Hao05}; Col.(7): optical spectral classifications of IR~QSOs (Seyfert~1 or QSO). The typical 
uncertainties of IR and optical luminosities are about 0.06 dex and 10-20\%, respectively \citep{Hao05}.\\
$^{a}$The 60 $\mu$m flux density is an upper limit.\\
\end{minipage}
\end{table*}

\clearpage

%
\begin{table*}
 \centering
 \begin{minipage}{140mm}
  \caption{ULIRG Sample}
  \begin{tabular}{@{}lcccccc@{}}
  \hline
~ & R.A. & Decl. & ~ & ~ & ~ & ~ \\
Object & (J2000.0) & (J2000.0) & Redshift & log$\left (L_{\rm 60\mu{\rm m}} / L_\odot \right)$ & 
$\alpha$(60, 25) & SpecType\\
~~(1) & (2) & (3) & (4) & (5) & (6) & (7)\\
  \hline  
F00188$-$0856 & 00~21~26.5 & $-$08~39~26 & 0.1284 & 12.166 & $-$2.22 & L\\
F00397$-$1312 & 00~42~15.5 & $-$12~56~03 & 0.2617 & 12.702 & $-$1.70$^{a}$ & H\\
F01004$-$2237 & 01~02~50.0 & $-$22~21~57 & 0.1177 & 12.031 & $-$1.42 & H\\ 
F01199$-$2307 & 01~22~20.9 & $-$22~52~07 & 0.1560 & 12.143 & $-$2.62$^{a}$ & H\\
F01298$-$0744 & 01~32~21.4 & $-$07~29~08 & 0.1362 & 12.201 & $-$2.49$^{a}$ & H\\
F01355$-$1814 & 01~37~57.4 & $-$17~59~21 & 0.1920 & 12.284 & $-$2.36$^{a}$ & H\\
Z03521+0028 & 03~54~42.2 & $+$00~37~03 & 0.1519 & 12.333 & $-$2.69 & L\\
F05189$-$2524 & 05~21~01.5 & $-$25~21~45 & 0.0426 & 11.880 & $-$1.57 & S2\\
F08572+3915 & 09~00~25.4 & $+$39~03~54 & 0.0584 & 11.899 & $-$1.68 & L\\
F09463+8141 & 09~53~00.5 & $+$81~27~28 & 0.1560 & 12.093 & $-$3.38$^{a}$ & L\\
F10091+4704 & 10~12~16.7 & $+$46~49~43 & 0.2460 & 12.451 & $-$3.02$^{a}$ & L\\
F10378+1108 & 10~40~29.2 & $+$10~53~18 & 0.1362 & 12.167 & $-$2.60 & L\\
F11095$-$0238 & 11~12~03.4 & $-$02~54~22 & 0.1066 & 12.091 & $-$2.34 & L\\
F11223$-$1244 & 11~24~50.0 & $-$13~01~13 & 0.1990 & 12.354 & $-$2.61$^{a}$ & S2\\
F11582+3020 & 12~00~46.8 & $+$30~04~15 & 0.2230 & 12.335 & $-$2.30$^{a}$ & L\\
F12032+1707 & 12~05~47.7 & $+$16~51~08 & 0.2170 & 12.389 & $-$2.01$^{a}$ & L\\
F12072$-$0444 & 12~09~45.1 & $-$05~01~14 & 0.1284 & 12.145 & $-$1.74 & S2\\
F12112+0305 & 12~13~46.0 & $+$02~48~38 & 0.0733 & 12.164 & $-$3.22 & L\\
Mrk~273 & 13~44~42.1 & $+$55~53~13 & 0.0378 & 11.974 & $-$2.57 & S2\\
F13451+1232 & 13~47~33.3 & $+$12~17~24 & 0.1217 & 11.985 & $-$1.20 & S2\\
F14070+0525 & 14~09~31.2 & $+$05~11~32 & 0.2644 & 12.611 & $-$2.15$^{a}$ & S2\\
F14348$-$1447 & 14~37~38.3 & $-$15~00~23 & 0.0827 & 12.182 & $-$3.00 & L\\
F15001+1433 & 15~02~31.9 & $+$14~21~35 & 0.1627 & 12.250 & $-$2.76 & S2\\
F15206+3342 & 15~22~38.0 & $+$33~31~36 & 0.1244 & 11.971 & $-$1.86 & H\\
Arp~220 & 15~34~57.1 & $+$23~30~11 & 0.0181 & 12.002 & $-$2.94 & H\\
F16090$-$0139 & 16~11~40.5 & $-$01~47~06 & 0.1336 & 12.348 & $-$2.99 & L\\
F16300+1558 & 16~32~21.4 & $+$15~51~45 & 0.2417 & 12.532 & $-$2.83$^{a}$ & L\\
F16333+4630 & 16~34~52.6 & $+$46~24~53 & 0.1910 & 12.207 & $-$2.87$^{a}$ & L\\
NGC~6240 & 16~52~58.9 & $+$02~24~03 & 0.0245 & 11.608 & $-$2.16 & L\\
F17068+4027 & 17~08~32.1 & $+$40~23~28 & 0.1790 & 12.192 & $-$2.73 & H\\
F17179+5444 & 17~18~54.2 & $+$54~41~47 & 0.1470 & 12.015 & $-$2.18 & S2\\
F20414$-$1651 & 20~44~18.2 & $-$16~40~16 & 0.0871 & 12.032 & $-$2.90 & H\\
F22491$-$1808 & 22~51~49.2 & $-$17~52~23 & 0.0778 & 12.024 & $-$2.62 & H\\
F23129+2548 & 23~15~21.4 & $+$26~04~32 & 0.1790 & 12.327 & $-$3.00$^{a}$ & L\\ 
F23498+2423 & 23~52~26.0 & $+$24~40~17 & 0.2120 & 12.244 & $-$1.82$^{a}$ & S2\\
\hline
\end{tabular}
~\\
~\\
Notes -- Basic properties of non-Seyfert 1 ULIRGs in our sample. Col.(1): source name; Col.(2)-(3): right ascension 
(hours, minutes, seconds) and declination (degrees, arcminutes, arcseconds), from NED; Col.(4): redshift, from NED; 
Col.(5): monochromatic luminosity at 60$\mu$m; Col.(6): infrared color index $\alpha$(60, 25); Col.(7): optical 
spectral type, derived from \citet{Veilleux99} and \citet{Wu98}. The typical uncertainty of IR luminosities is about 
0.06 dex \citep{Hao05}.\\
$^{a}$The 25 $\mu$m flux density is an upper limit.\\
\end{minipage}
\end{table*}

%
\begin{table*}
 \centering
 \begin{minipage}{140mm}
  \caption{IRS Observation Log of IR~QSOs}
  \begin{tabular}{@{}lcccccc@{}}
  \hline
Object & SL & LL & PID (L) & SH & LH & PID (H)\\
~~(1) & (2) & (3) & (4) & (5) & (6) & (7)\\
\hline
I~Zw~1 & 2$\times$14.68 & 2$\times$14.68 & 14 & 4$\times$31.46 & 2$\times$60.95 & 14\\
F00275$-$2859 & 2$\times$60.95 & 2$\times$31.46 & 105 & 3$\times$121.9 & 8$\times$60.95 & 1096\\
3C~48 & 1$\times$6.29 & 1$\times$14.68 & 82 & \ldots & \ldots & \ldots\\
Mrk~1014 & 3$\times$14.68 & 2$\times$31.46 & 105 & 6$\times$31.46 & 4$\times$60.95 & 105\\
F02054+0835 & 2$\times$60.95 & 2$\times$31.46 & 105 & \ldots & \ldots & \ldots\\
F07599+6508 & 3$\times$14.68 & 2$\times$31.46 & 105 & 6$\times$31.46 & 4$\times$60.95 & 105\\
F11119+3257 & 1$\times$60.95 & 3$\times$31.46 & 105 & 3$\times$121.9 & 4$\times$60.95 & 105\\
3C~273 & 3$\times$14.68 & 3$\times$14.68 & 105 & 6$\times$31.46 & 4$\times$60.95 & 105\\
Mrk~231 & 2$\times$14.68 & 5$\times$6.29 & 105 & 6$\times$31.46 & 4$\times$60.95 & 105\\
F13218+0552 & 1$\times$60.95 & 3$\times$31.46 & 105 & 3$\times$121.9 & 4$\times$60.95 & 105\\
F13342+3932 & 2$\times$60.95 & 2$\times$31.46 & 105 & 3$\times$121.9 & 2$\times$241.83 & 105\\
F14026+4341 & 7$\times$14.68 & 7$\times$14.68 & 61 & 14$\times$121.9 & 28$\times$60.95 & 61\\
PG~1543+489 & 2$\times$60.95 & 2$\times$121.9 & 20142 & \ldots & \ldots & \ldots\\
F15462$-$0450 & 1$\times$60.95 & 3$\times$31.46 & 105 & 3$\times$121.9 & 4$\times$60.95 & 105\\
PG~1613+658 & 2$\times$60.95 & 2$\times$121.9 & 3187 \& 20142 & 3$\times$121.9 & 10$\times$241.83 & 3187\\
PG~1700+518 & 1$\times$6.29 & 1$\times$14.68 & 82 & 2$\times$121.9 & 7$\times$241.83 & 3187\\
F18216+6419 & 1$\times$14.68 & 1$\times$31.46 & 82 & 3$\times$121.9 & 3$\times$241.83 & 3746\\
F20036$-$1547 & 2$\times$60.95 & 2$\times$31.46 & 105 & \ldots & \ldots & \ldots\\
F21219$-$1757 & 2$\times$60.95 & \ldots & 3187 & 2$\times$121.9 & 2$\times$241.83 & 3187\\
\hline
\end{tabular}
~\\
~\\
Notes -- The integration times for each slit are given. The quantity is the number of DCEs ``$\times$'' 
ramp time (in sec) for a single nod. Col.(1): source name; Col.(2)-(4): integration times and program IDs for 
the low-resolution (short-low [SL] \& long-low [LL] modes) observations; Col.(5)-(7): integration times and 
program IDs for the high-resolution (short-high [SH] \& long-high [LH] modes) observations.\\
\end{minipage}
\end{table*}

\clearpage

%
\begin{table*}
 \centering
 \begin{minipage}{140mm}
  \caption{IRS Observation Log of PG~QSOs and ULIRGs}
  \begin{tabular}{@{}lcccc@{}}
  \hline
Object & SL & PID (SL) & LL & PID (LL)\\
~~(1) & (2) & (3) & (4) & (5)\\
  \hline
\multicolumn{5}{c}{PG~QSOs}\\
  \hline
PG~0804+761 & 3$\times$14.68 & 14 & 2$\times$31.46 & 14\\
PG~0838+770 & 2$\times$241.83 & 3187 & 2$\times$121.9 & 20142\\
PG~0844+349 & 2$\times$60.95 & 3187 & 2$\times$121.9 & 20142\\
PG~1004+130 & 2$\times$60.95 & 20142 & 2$\times$121.9 & 20142\\
PG~1119+120 & 3$\times$14.68 & 14 & 2$\times$31.46 & 14\\
PG~1149$-$110 & 2$\times$60.95 & 20142 & 2$\times$121.9 & 20142\\
PG~1211+143 & 3$\times$14.68 & 14 & 2$\times$31.46 & 14\\
PG~1351+640 & 3$\times$14.68 & 14 & 2$\times$31.46 & 14\\
PG~1411+442 & 2$\times$60.95 & 3187 & 1$\times$121.9 & 3421\\
PG~1426+015 & 2$\times$60.95 & 3187 & 2$\times$121.9 & 20142\\
PG~1444+407 & 2$\times$60.95 & 20142 & 2$\times$121.9 & 20142\\
PG~1501+106 & 3$\times$14.68 & 14 & 2$\times$31.46 & 14\\
PG~1519+226 & 2$\times$60.95 & 20142 & 2$\times$121.9 & 20142\\
PG~1534+580 & 2$\times$60.95 & 20142 & 2$\times$121.9 & 20142\\
PG~1535+547 & 3$\times$60.95 & 3421 & 1$\times$121.9 & 3421\\
PG~1552+085 & 2$\times$60.95 & 20142 & 2$\times$121.9 & 20142\\
PG~1612+261 & 2$\times$60.95 & 20142 & 2$\times$121.9 & 20142\\
PG~1617+175 & 3$\times$60.95 & 3187 & 2$\times$121.9 & 20142\\
PG~2130+099 & 3$\times$14.68 & 14 & 2$\times$31.46 & 14\\
  \hline
\multicolumn{5}{c}{ULIRGs}\\
  \hline
F00188$-$0856 & 2$\times$60.95 & 105 & 3$\times$31.46 & 105\\
F00397$-$1312 & 2$\times$60.95 & 105 & 3$\times$31.46 & 105\\
F01004$-$2237 & 1$\times$60.95 & 105 & 2$\times$31.46 & 105\\
F01199$-$2307 & 2$\times$60.95 & 105 & 3$\times$31.46 & 105\\
F01298$-$0744 & 2$\times$60.95 & 105 & 2$\times$31.46 & 105\\
F01355$-$1814 & 2$\times$60.95 & 105 & 2$\times$31.46 & 105\\
Z03521+0028 & 2$\times$60.95 & 105 & 3$\times$31.46 & 105\\
F05189$-$2524 & 3$\times$14.68 & 105 & 2$\times$14.68 & 105\\
F08572+3915 & 3$\times$14.68 & 105 & 3$\times$14.68 & 105\\
F09463+8141 & 2$\times$60.95 & 105 & 2$\times$31.46 & 105\\
F10091+4704 & 2$\times$31.46 & 105 & 2$\times$31.46 & 105\\
F10378+1108 & 2$\times$60.95 & 105 & 3$\times$31.46 & 105\\
F11095$-$0238 & 2$\times$60.95 & 105 & 2$\times$31.46 & 105\\
F11223$-$1244 & 2$\times$60.95 & 105 & 3$\times$31.46 & 105\\
F11582+3020 & 2$\times$60.95 & 105 & 2$\times$31.46 & 105\\
F12032+1707 & 2$\times$60.95 & 105 & 2$\times$31.46 & 105\\
F12072$-$0444 & 1$\times$60.95 & 105 & 2$\times$31.46 & 105\\
F12112+0305 & 3$\times$14.68 & 105 & 2$\times$31.46 & 105\\
Mrk~273 & 2$\times$14.68 & 105 & 2$\times$14.68 & 105\\
F13451+1232 & 3$\times$14.68 & 105 & 2$\times$31.46 & 105\\
F14070+0525 & 2$\times$60.95 & 105 & 2$\times$31.46 & 105\\
F14348$-$1447 & 1$\times$60.95 & 105 & 2$\times$31.46 & 105\\
F15001+1433 & 2$\times$60.95 & 105 & 3$\times$31.46 & 105\\
F15206+3342 & 1$\times$60.95 & 105 & 3$\times$31.46 & 105\\
Arp~220 & 3$\times$14.68 & 105 & 5$\times$6.29 & 105\\
F16090$-$0139 & 1$\times$60.95 & 105 & 3$\times$31.46 & 105\\
F16300+1558 & 2$\times$60.95 & 105 & 5$\times$31.46 & 105\\
F16333+4630 & 2$\times$60.95 & 105 & 2$\times$31.46 & 105\\
NGC~6240 & 2$\times$14.68 & 105 & 2$\times$14.68 & 105\\
F17068+4027 & 2$\times$60.95 & 105 & 3$\times$31.46 & 105\\
F17179+5444 & 2$\times$60.95 & 105 & 3$\times$31.46 & 105\\
F20414$-$1651 & 1$\times$60.95 & 105 & 2$\times$31.46 & 105\\
F22491$-$1808 & 1$\times$60.95 & 105 & 2$\times$31.46 & 105\\
F23129+2548 & 3$\times$60.95 & 105 & 5$\times$31.46 & 105\\
F23498+2423 & 2$\times$60.95 & 105 & 2$\times$31.46 & 105\\
\hline
\end{tabular}
~\\
~\\
Notes -- The integration times for each slit are given. The quantity is the number of DCEs ``$\times$'' 
ramp time (in sec) for a single nod. Col.(1): source name; Col.(2)-(3): integration times and program IDs 
for the short-low (SL) mode observations; Col.(4)-(5): integration times and program IDs for the long-low 
(LL) mode observations.\\
\end{minipage}
\end{table*}

%
\begin{table*}
 \centering
 \begin{minipage}{140mm}
  \caption{Mid-IR properties of IR~QSOs}
  \begin{tabular}{@{}lccccc@{}}
  \hline
~ & S$_{15}$ & S$_{30}$ & PAH 6.2$\mu$m & EW(PAH) & [NeII]12.81$\mu$m\\
Object & (mJy) & (mJy) & (10$^{-21}$W cm$^{-2}$) & ($\mu$m) & (10$^{-21}$W cm$^{-2}$)\\
~~(1) & (2) & (3) & (4) & (5) & (6)\\
  \hline
I~Zw~1 & 553.6 & 1186.0 & 17.2 & 0.011 & 2.87\\
F00275$-$2859 & 93.9 & 249.7 & 7.36 & 0.028 & 1.29\\
3C~48 & 75.6 & 283.8$^{a}$ & $<$7.2 & $<$0.051 & \ldots\\
Mrk~1014 & 225.9 & 1191.3 & 12.9 & 0.048 & 5.94\\
F02054+0835 & 127.8 & 260.6$^{a}$ & $<$4.0 & $<$0.018 & \ldots\\
F07599+6508 & 285.2 & 754.9 & 5.85 & 0.004 & 3.29\\
F11119+3257 & 188.1 & 645.6 & 7.52 & 0.014 & 2.83\\
3C~273 & 336.7 & 410.4 & $<$9.4 & $<$0.004 & 1.26\\
Mrk~231 & 3385.6 & 13241.2 & 81.8 & 0.012 & 18.24\\
F13218+0552 & 288.1 & 734.2 & 6.80 & 0.005 & 1.34\\
F13342+3932 & 101.7 & 373.1 & 16.1 & 0.089 & 5.26\\
F14026+4341 & 140.0 & 267.0$^{a}$ & 4.45 & 0.012 & 0.84\\
PG~1543+489 & 69.2 & 125.3$^{a}$ & $<$6.1 & $<$0.034 & \ldots\\ 
F15462$-$0450 & 140.7 & 913.8 & 23.2 & 0.099 & 8.13\\
PG~1613+658 & 127.1 & 261.0 & 7.80 & 0.020 & 4.07\\
PG~1700+518 & 122.8 & 246.6$^{a}$ & $<$9.5 & $<$0.022 & 2.09\\
F18216+6419 & 240.8 & 365.4$^{a}$ & $<$8.1 & $<$0.015 & 3.82\\
F20036$-$1547 & 95.9 & 441.4 & 12.0 & 0.046 & \ldots\\
F21219$-$1757 & \ldots & \ldots & 8.47 & 0.018 & 1.71\\
\hline
\end{tabular}
~\\
~\\
Notes -- Mid-IR spectral properties of IR~QSOs. Col.(1): source name; Col.(2)-(3): Continuum flux densities 
around 15 and 30$\mu$m rest wavelength, the uncertainties of the measurements are typically 10$\%$; Col.(4)-(5): 
Flux and equivalent width (EW) of PAH emissions at 6.2$\mu$m, upper limits (3$\sigma$) are derived adopting 
widths of 0.2$\mu$m; Col.(6) fluxes of [NeII] 12.81$\mu$m line.\\
$^{a}$Outside the rest wavelength range available, the value given here was linearly extrapolated based on 
the logarithm of the $\sim$25$\mu$m rest wavelength continuum level.\\
\end{minipage}
\end{table*}

\clearpage

%
\begin{table*}
 \centering
 \begin{minipage}{140mm}
  \caption{Mid-IR properties of PG~QSOs and ULIRGs}
  \begin{tabular}{@{}lcccc@{}}
  \hline
~ & S$_{15}$ & S$_{30}$ & PAH 6.2$\mu$m & EW(PAH)\\
Object & (mJy) & (mJy) & (10$^{-21}$W cm$^{-2}$) & ($\mu$m)\\
~~(1) & (2) & (3) & (4) & (5)\\
  \hline
\multicolumn{5}{c}{PG~QSOs}\\
  \hline
PG~0804+761 & 125.5 & 150.8 & $<$7.4 & $<$0.012\\
PG~0838+770 & 45.4 & 97.4 & $<$2.0 & $<$0.026\\
PG~0844+349 & 105.5 & 155.1 & $<$6.5 & $<$0.018\\
PG~1004+130 & 72.9 & 118.5 & $<$2.0 & $<$0.014\\
PG~1119+120 & 176.9 & 303.9 & $<$2.1 & $<$0.008\\
PG~1149$-$110 & 140.3 & 249.6 & $<$2.1 & $<$0.019\\
PG~1211+143 & 239.6 & 268.8 & $<$7.5 & $<$0.014\\
PG~1351+640 & 249.4 & 536.2 & 10.9 & 0.034\\
PG~1411+442 & 125.3 & 143.2 & $<$4.2 & $<$0.010\\
PG~1426+015 & 136.1 & 181.2 & $<$5.2 & $<$0.013\\
PG~1444+407 & 69.8 & 99.1 & $<$7.54 & $<$0.036\\
PG~1501+106 & 297.6 & 424.9 & $<$3.4 & $<$0.008\\
PG~1519+226 & 54.2 & 70.8 & 2.46 & 0.016\\
PG~1534+580 & 107.2 & 149.1 & $<$2.08 & $<$0.008\\
PG~1535+547 & 98.4 & 111.5 & $<$3.1 & $<$0.011\\
PG~1552+085 & 79.7 & 94.2 & $<$1.3 & $<$0.017\\
PG~1612+261 & 83.4 & 136.4 & $<$3.0 & $<$0.020\\
PG~1617+175 & 53.8 & 65.7 & $<$3.2 & $<$0.020\\
PG~2130+099 & 196.4 & 293.0 & 3.56 & 0.006\\
  \hline
\multicolumn{5}{c}{ULIRGs}\\
  \hline
F00188$-$0856 & 122.8 & 1052.7 & 27.5 & 0.187\\
F00397$-$1312 & 90.1 & 523.5 & 9.14 & 0.018\\
F01004$-$2237 & 306.4 & 1176.4 & 11.2 & 0.048\\
F01199$-$2307 & 39.1 & 455.0 & 3.75 & 0.146\\
F01298$-$0744 & 65.4 & 718.5 & 7.53 & 0.092\\
F01355$-$1814 & 27.7 & 369.6 & $<$2.7 & $<$0.186\\
Z03521+0028 & 50.6 & 980.1 & 20.9 & 1.086\\
F05189$-$2524 & 1017.6 & 4994.1 & 54.8 & 0.034\\
F08572+3915 & 657.6 & 3272.9 & $<$7.7 & $<$0.004\\
F09463+8141 & 9.0 & 193.3 & 7.61 & 1.483\\
F10091+4704 & 13.3 & 197.1 & 7.06 & 0.758\\
F10378+1108 & 48.5 & 571.9 & 11.0 & 0.415\\
F11095$-$0238 & 111.8 & 1112.0 & $<$3.7 & $<$0.025\\
F11223$-$1244 & 19.0 & 214.3 & $<$2.2 & $<$0.065\\
F11582+3020 & 32.5 & 352.8 & 1.25 & 0.022\\
F12032+1707 & 45.5 & 428.8 & $<$7.1 & $<$0.161\\
F12072$-$0444 & 197.3 & 878.6 & 13.6 & 0.106\\
F12112+0305 & 106.4 & 1683.5 & 56.8 & 0.431\\ 
Mrk~273 & 416.1 & 5068.8 & 74.8 & 0.129\\
F13451+1232 & 286.5 & 996.7 & $<$3.0 & $<$0.013\\
F14070+0525 & 21.6 & 304.0 & 5.59 & 0.187\\
F14348$-$1447 & 82.7 & 1309.2 & 33.5 & 0.365\\
F15001+1433 & 59.6 & 475.1 & 16.1 & 0.184\\
F15206+3342 & 101.6 & 628.9 & 38.2 & 0.342\\
Arp~220 & 1048.1 & 20907.7 & 161.0 & 0.272\\
F16090$-$0139 & 76.5 & 808.3 & 13.5 & 0.107\\
F16300+1558 & 25.7 & 490.4 & 11.0 & 0.282\\
F16333+4630 & 20.2 & 260.9 & 12.5 & 0.529\\
NGC~6240 & 752.1 & 5763.0 & 313.0 & 0.365\\ 
F17068+4027 & 52.3 & 410.9 & 10.8 & 0.198\\
F17179+5444 & 72.6 & 341.8 & 6.43 & 0.099\\
F20414$-$1651 & 61.7 & 1397.4 & 37.5 & 0.587\\
F22491$-$1808 & 82.9 & 1473.3 & 23.1 & 0.465\\
F23129+2548 & 40.2 & 434.9 & $<$1.1 & $<$0.021\\
F23498+2423 & 46.4 & 239.7 & 6.11 & 0.053\\
\hline
\end{tabular}
~\\
~\\
Notes -- Mid-IR spectral properties of PG~QSOs and ULIRGs in our sample. The definition of columns 
is the same as in Table~5.\\
\end{minipage}
\end{table*}

\label{lastpage}

\end{document}